\documentclass[5p,preprint,numbers,sort&compress]{elsarticle}
\usepackage{lineno}
\usepackage[utf8]{inputenc}
\usepackage{geometry}
\usepackage{pdfpages}
\usepackage{amsmath}
\usepackage{amsthm}
\usepackage{amssymb}
\usepackage{amsfonts}
\usepackage{miller}
\usepackage{graphicx}
\usepackage[dvipsnames,table]{xcolor}
\usepackage{textcomp}
\usepackage{nicefrac}
\usepackage{siunitx}
\usepackage{comment}
\usepackage{booktabs}      
\usepackage{multicol}      
\usepackage{multirow}      
\usepackage{array}         
\usepackage[outercaption]{sidecap}
\usepackage{boldline}
\usepackage{stackengine}  %
\usepackage{tikz}
\usetikzlibrary{external}
\def\degree{$^{\circ}$}
\usepackage[T1]{fontenc}
\usepackage[utf8]{inputenc}
\usepackage{placeins}
\usepackage[normalem]{ulem}

\usepackage{subcaption}
\usepackage{caption}
\usepackage{listing}   
\usepackage{hyperref}
\usepackage{enumitem}
\usepackage{tabularx}
\usepackage{cuted}

\DeclareFontFamily{U}{cry}{\hyphenchar\font=-1}
\DeclareFontShape{U}{cry}{m}{n}{ <-> cryst}{}

\usepackage{geometry}
\begin{document}
\sloppy

\title{Resolving Overlapping EBSD Patterns by Experiment - Simulation Residuals Analysis}

\author[label1]{Grzegorz Cios\corref{cor1}}
\address[label1]{Academic Centre for Materials and Nanotechnology, AGH University of Krakow, al.\@ A. Mickiewicza 30, 30-059 Krakow, Poland}
\ead{Ciosu@agh.edu.pl}
\cortext[cor1]{Corresponding author}

\author[label1]{Aimo Winkelmann}

\author[label1]{Tomasz Tokarski}
\author[label3]{Wiktor Bednarczyk}
\author[label3,label1]{Piotr Bała}
\address[label3]{Faculty of Metals and Industrial Computer Science, AGH University of Krakow, al.\@ A. Mickiewicza 30, 30-059 Krakow, Poland}


\begin{abstract}
In the technique of Electron Backscatter Diffraction (EBSD), the accurate detection and identification of different phases existing in a sample is often limited by overlapping Kikuchi diffraction patterns originating from the extended probing volume of the individual EBSD map points measured in the scanning electron microscope (SEM). 
We present an iterative approach that uses simulated Kikuchi patterns to resolve several overlapping diffraction signals. 
For each measured EBSD pattern, our method first identifies the best-fit simulated Kikuchi pattern using dynamic template matching. 
This simulated, ideal reference pattern is then further processed to optimally match the experimental image, uncovering any underlying weaker signals after subtraction. 
Repeatedly utilizing dynamic template matching and pattern subtraction on residual signals of subsequent steps enables the identification of minor phases that might otherwise be missed from the probing volume of the EBSD map point.
This method significantly improves phase detection in complex materials, addressing a key limitation of conventional EBSD analysis that conventionally assigns a single phase to each map point. 
The present method does not require a known orientation relationship between the phases of the overlapping patterns or close neighbor experimental patterns like previously published approaches.
\end{abstract}



\maketitle



\section{Introduction} \label{sec:Intro}
Electron backscatter diffraction is a powerful and widely used technique for the crystallographic characterization of materials. 
EBSD enables us to map microstructural features such as grain orientations, boundaries, and phase distributions with sub-micrometer spatial resolution. 

Across various EBSD applications, several spatial resolution definitions are used in the literature  \cite{Zaefferer2007,Chen2011}. Here, the emphasis is on the \textit{physical resolution} as the most meaningful measure: it can be defined by the lateral distance from a true high‑angle grain boundary over which the raw Kikuchi signal remains a superposition from both crystals, and it is fundamentally governed by the SEM–sample interaction volume. 
In contrast, the so‑called \mbox{\textit{effective resolution}} is an artifact of the acquisition-analysis workflow, because, under the prevailing EBSD visualization paradigm that enforces a single phase and a single orientation at every map point, software algorithms must place a well-defined boundary within a region containing mixed contributions. 
As a result, the reported “effective” resolution depends on user conventions and indexing heuristics rather than on physics, and it can appear numerically finer than the physical limit simply because of how grain boundaries are defined and classified for visualization in an EBSD map. 
It is understood that actual grain boundaries, which are defined on an atomic level, can never be imaged by EBSD because the required spatial resolution is only available in TEMs.

Related to the different size and shape of the three-dimensional SEM beam interaction volume, the reported spatial resolutions depend on the sample material and scanning directions.
On polycrystalline Fe, Zaefferer~\cite{Zaefferer2007} reported a resolution of 30~× 90 nm , depending on the directions parallel and perpendicular to the SEM tilt axis. 
In contrast, Tripathi~\cite{Tripathi2019} showed that for magnesium, the resolution can vary from 240 nm at 5 kV to about 575 nm at 15 kV. 
These values highlight that pattern overlap from different phases and crystal orientations in the extended sampling volume of the SEM can often be the limiting factor for true physical resolution of EBSD data.

Steinmetz and Zaefferer~\cite{Steinmetz2010} demonstrated that lowering the accelerating voltage reduces the volume of the electron interaction and therefore the pattern overlap, improving the effective spatial resolution from about 30 nm at 15 kV to 10 nm at 7.5 kV. Their approach used changes in Hough peak intensity across a boundary as a more quantitative measure of overlap effects, but they also noted that lower voltages demand longer acquisition times, increasing risks of drift and contamination.

Bate et al.~\cite{BATE2005} showed that the effective sampling area in EBSD is often much larger than the nominal beam size, so overlapping patterns at low-angle grain boundaries can create artifacts that mimic gradual orientation changes. Although advanced pattern-shift methods can significantly improve angular resolution, intrinsic pattern overlap near boundaries remains a limiting factor for resolving fine details.

Wright et al.~\cite{Wright2014} also demonstrated that overlapping EBSD patterns at grain boundaries — due to pattern mixing and noise from high dislocation densities — can produce indexing artifacts, typically introducing errors around 0.5\degree{}. Such artifacts complicate the interpretation of local misorientations, but can be partly mitigated by using more Kikuchi bands and careful offline indexing.

Shi et al.~\cite{Shi2021} proposed an extended DIC-based EBSD indexing method for overlapped EBSPs, capable of extracting multiple crystal orientations with high precision. Although this improves resolution at grain boundaries, it requires high-quality data, assumes global pattern matching, and can be computationally intensive.

Tong et al.~\cite{Tong2015} quantified how pattern overlap affects the precision of high-resolution cross-correlation EBSD (HR-EBSD), showing that mixing within 18 nm of a boundary can degrade strain measurements. They demonstrated that while careful Fourier filtering can help, over-filtering or ignoring overlap can introduce significant errors.

However, the effective resolution of EBSD is ultimately limited by the volume of electron-sample interaction, which can extend tens to hundreds of nanometers depending on the beam parameters and the density of the material~\cite{Zaefferer2007, Steinmetz2010, Chen2011}. In fine-grained, multiphase, or nanostructured materials, EBSD patterns often contain Kikuchi signals that overlap from multiple crystals or phases within this volume~\cite{Wright2014, Lenthe2020}. 
Traditional Hough-transform-based indexing typically assumes a single dominant orientation per pattern and struggles with overlapping signals, leading to misindexing or failed solutions~\cite{Shen2021, Chen2011}.

Recent advances in full-pattern matching and dictionary-based indexing (DI) offer better handling of noisy or overlapped patterns~\cite{Shi2021, Lenthe2020, Nolze2018}, but require high-quality patterns and significant computing resources. 
In particular, Brodu et al.\cite{Brodu2022} showed that weak secondary Kikuchi signals from nanoscale phases can be recovered by subtracting a high-quality neighboring pattern, enabling, for example, the mapping of sub-100 nm cementite lamellae in nanostructured pearlitic steel at 15~kV.

These developments demonstrate that the conventional resolution limit, dominated by interaction volume size, can be extended by advanced computational post-processing — a concept sometimes described as the effective spatial resolution. Hough-based indexing typically resolves a stronger dominant signal in an overlapping pattern. For example, Chauniyal et al.~\cite{Chauniyal2024} used constrained non-negative matrix factorization to decompose the overlapped signals into their contributing orientations and quantify each contribution.

Despite these advances, challenges remain: separating signals in multigrain overlaps is still difficult, long exposure times increase drift and contamination risk, and post-processing can be time-consuming and parameter-sensitive. Continued progress in both hardware and algorithms will be essential to push EBSD toward routine, nanoscale crystallographic mapping of complex materials.

In contrast to existing approaches, the present work introduces a fully simulation-driven residual analysis that does not require prior knowledge of orientation relationships or access to non-overlapping reference patterns. By iteratively subtracting dynamically fitted simulated EBSD patterns, multiple crystallographic contributions can be resolved within a single EBSD map pixel.

\section{Materials and Methods}

\subsection{Materials}
The materials investigated were:
\begin{itemize}
  \item annealed C45 ferritic--pearlitic steel ($\sim$0.45~wt.\%~C);
  \item Cu--Al--Ni--Fe two-phase bronze (FCC + BCC);
  \item dual-phase steel obtained by quenching from 740~$^\circ$C;
  \item AISI~H11 tool steel with spheroidized Cr$_7$C$_3$ carbides;
  \item CuPt thin film deposited on a SrTiO3 (STO) (111) single-crystal.
\end{itemize}

\begin{table*}
    \centering
    \begin{tabular}{|c|c|c|c|c|c|c|}
        \hline
        \textbf{Sample} & \textbf{Phases} & \textbf{State} & \textbf{Step Size} & \textbf{Acq. Speed} & \textbf{Frame ave.} \\
        \hline
        C45 steel & $\alpha$-Fe, Fe$_3$C & annealed & 20 nm  & 446 pps & $2\times$ \\
        \hline
        two-phase bronze & Cu$_{FCC}$, BCC & as-built & 25 nm & 1508 pps & $2\times$ \\
        \hline
        dual-phase steel & $\alpha$-Fe, $\gamma$-Fe & as quenched & 200 nm & 228 pps & $4\times$ \\
        \hline
        AISI~H11 & $\alpha$-Fe, Cr$_7$C$_3$ & annealed & 100 nm & 1000 pps& $1\times$\\ 
        \hline
        CuPt thin film & Cu$_{FCC}$ & as deposited & 500 nm & 749 pps & $1\times$\\
        \hline
    \end{tabular}
    \caption{Summary of experimental conditions for the different samples (pattern resolution was 156$\times$128 in each case)}
    \label{tab:samples}
\end{table*}

\subsection{EBSD}

The EBSD setup applied in this study consists of a Helios pFIB field emission gun SEM (Thermo Fisher Scientific), which was equipped with a Symmetry S3 EBSD detector (Oxford Instruments Nanoanalysis). For further analysis, all diffraction patterns were stored in a single file using AZtec v. 6.1 (Oxford Instruments Nanoanalysis) and subsequently transferred to a file format compatible with HDF5 \cite{hdf5} of type \texttt{.h5oina} \cite{Pinard2021}. 
For all samples the microscope was operated at 20 kV and 26 nA, the EBSD detector was set to Speed 2 mode with $156 \times 128$ pixel resolution.
The data acquisition parameters for each sample are summarized in table \ref{tab:samples}. For Dynamic Template Matching (DTM), the commercial software MapSweeper (AZtecCrystal v. 3.3 SP1) was used. The required physics-based master pattern has been simulated in AZtecCrystal as well. The dynamical simulations are based on the same approach as described in \cite{winkelmann2007um}.

\subsection{Residual signal and gain map fitting}

After orientation refinement in AztecCrystal MapSweeper, the best-fit simulated pattern was further post-processed to match the experimental pattern as closely as possible before subtraction. 
The post-processing involved two steps: first, the simulated pattern was blurred using a Gaussian kernel with standard deviation \( \sigma \), and second, a pixel-wise gain correction mask was applied to account for the reduced signal-to-noise ratio in the outer parts of the detected patterns. 
The gain mask, operating on the zero-mean signal, was described by a power law function of the normalized elliptical radius:
\begin{equation}
    G(r_{\mathrm{ell}}) = g_{\min} + (1 - r_{\mathrm{ell}})^{p}\,(g_{\max} - g_{\min}),
\end{equation}
where \( r_{\mathrm{ell}} \) is the normalized elliptical radius with respect to the image center, and \( g_{\min} \), \( g_{\max} \), and \( p \) are the fitted parameters. Alternatively, to reduce computational cost, the center coordinates of the ellipse (\( x_0 \), \( y_0 \)) were estimated by calculating the center of mass of the normalized cross-correlation (NCC) map between the experimental and simulated patterns. For this purpose, both patterns were divided into a grid of tiles \( 5 \times 5 \), and the NCC was calculated separately within each tile. The resulting \( 5 \times 5 \) NCC map was then used to determine the center of mass, which was taken as the estimated ellipse center.

All these parameters, together with the Gaussian blur width \( \sigma \), were included in the optimization.

All model parameters were determined by minimizing the weighted sum of squared residuals (SSR) between the normalized experimental pattern \( E(x, y) \) and the processed simulated pattern:
\begin{align}
    S_{G}(x, y) &= (S * K_{\sigma})(x, y) \\
    S'(x, y) &= S_{G}(x, y) \cdot G(x, y) \\
    \mathrm{Residual}(x, y) &= E(x, y) - \mathrm{NCC}(E, S') \cdot S'(x, y) \\
    \mathrm{SSR} &= \sum_{x, y}\, \mathrm{Residual}(x, y)^2
\end{align}
where \( S(x, y) \) is the simulated pattern, \( K_{\sigma} \) is the Gaussian kernel, \( G(x, y) \) is the gain map and \( \mathrm{NCC}(E, S') \) denotes the normalized cross-correlation coefficient between the experimental pattern and the simulated pattern corrected for gain with Gaussian stripes. Before fitting, both \( E \) and \( S' \) were normalized to zero mean and unit variance.

The choice to scale the simulated pattern \( S'(x, y) \) by \( \mathrm{NCC}(E, S') \) prior to subtraction arises from the fact that this value represents the optimal scalar \( \alpha^* \) that minimizes the residual norm (see Appendix):
\begin{equation}
    \alpha^* = \arg\min_\alpha \|E - \alpha S'\|^2 = \frac{\langle E, S' \rangle}{\|S'\|^2}
\end{equation}
Given the normalization \( \|E\| = \|S'\| = 1 \), this simplifies to:
\begin{equation}
    \alpha^* = \langle E, S' \rangle = \mathrm{NCC}(E, S')
\end{equation}
Thus, the scaling factor \( \mathrm{NCC}(E, S') \) ensures that the residual
\begin{equation}
\mathrm{Residual}(x, y) = E(x, y) - \mathrm{NCC}(E, S') \cdot S'(x, y)
\end{equation}
is orthogonal to \( S'(x, y) \) and has the minimum possible intensity. This projection effectively removes the best-matching component from the simulation, and the resulting residual captures only the uncorrelated features, typically appearing structurally flat or "gray". 

\begin{figure}[htb!]
    \centering
    \includegraphics[width=\columnwidth,clip]{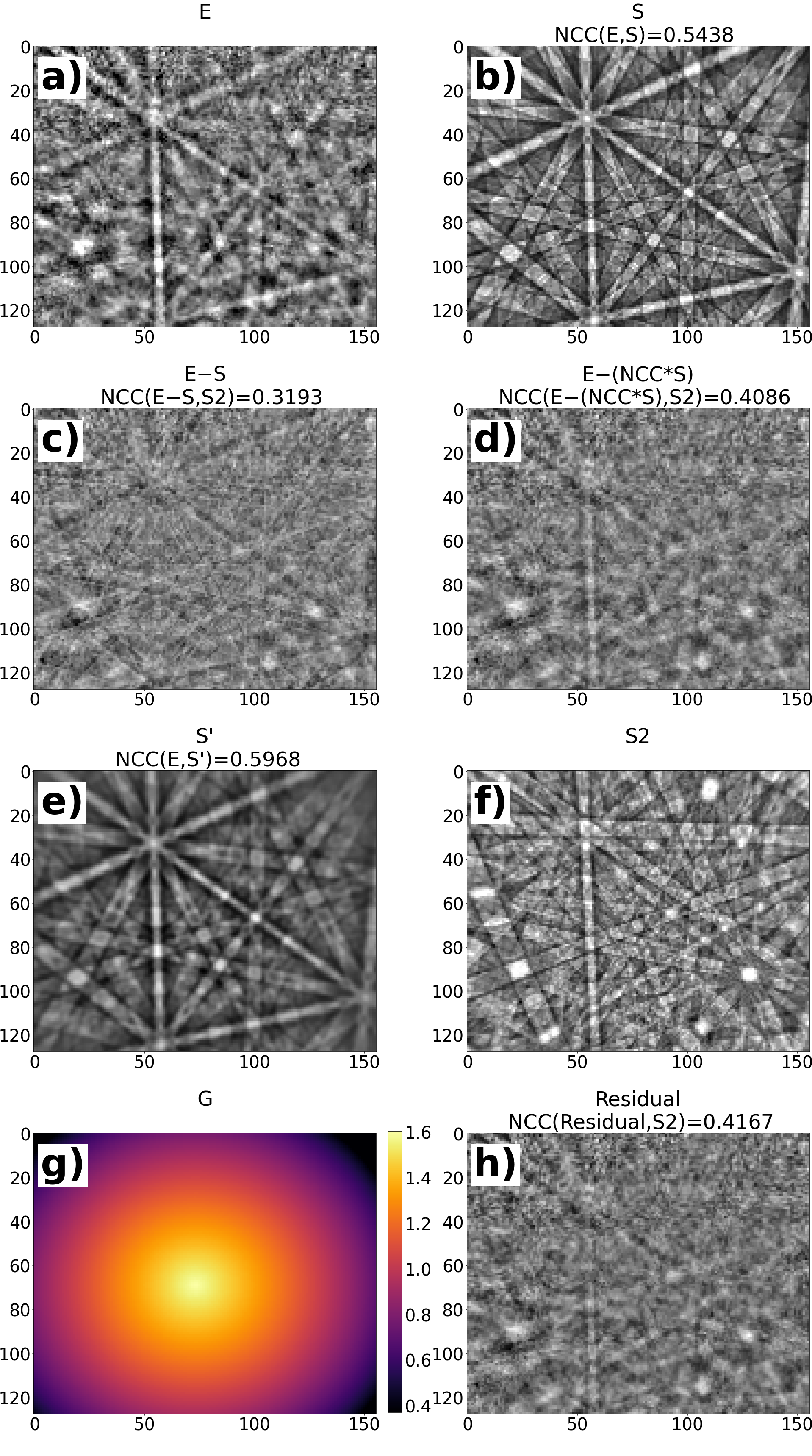}
    \caption{Residual patterns generated using different methods, a) Experimental pattern from cementite lamella in pearlite, b) Best fit of BCC simulation to a), c) Difference of a) and b), d) Difference of a) and b)*NCC(E,S), e) Best fit simulation blurred and multiplied by gain mask, f) Best fit simulation for overlapping cementite pattern, g) Fitted gain mask, h) Residual calculated according to Eq. 8.}
    \label{fig:Residual_pat}
\end{figure}

Different approaches of residual generation are shown in Fig.\ref{fig:Residual_pat}. The standard simulation shown in b) has NCC=0.5438 and after subtraction from the experiment the residual pattern in c) has clearly visible dark bands which suggests that in these areas the pattern was subtracted too much. 
To overcome this problem using the reasoning of Eq. 6 and 7 the simulated pattern was multiplied by NCC before subtraction in d). This allowed to minimize the residual norm. In other words, the amount of subtracted intensity was scaled to fit the experimental intensities, reducing the magnitude of dark bands visible in c). Fitting an additional Gaussian blur and multiplying by the gain mask (g) allows to increase the NCC of the simulated image from 0.5438 to 0.5968 in e). 
The best fit cementite pattern is shown in Fig.\ref{fig:Residual_pat}(f) with the NCCs being 0.3193 for standard subtraction (Fig.\ref{fig:Residual_pat}(c)), 0.4086 for NCC weighted subtraction (Fig.\ref{fig:Residual_pat}(d)) and 0.4167 for the complete processing with Gaussian blur, gain masking and NCC weighting as shown in Fig.\ref{fig:Residual_pat}(h). 
Additional representative examples of experimental EBSPs, corresponding best-fit simulations, and residual patterns obtained using this procedure are provided in the Supplementary Information (Figs. S1–S10).

The final residual patterns were stored in the corresponding \texttt{.h5oina} file for subsequent analysis, including DTM in the MapSweeper module of AZtec Crystal (v. 3.3). Based on empirical observations, residual patterns with NCC values below approximately 0.175 were found to be dominated by noise and were therefore excluded from further analysis.

\section{Results and Discussion}

\subsection*{Ferritic-pearlitic steel}
\begin{figure}[htb!]
\begin{center}
    \begin{subfigure}{.5\textwidth}
         \includegraphics[clip,width=\textwidth]{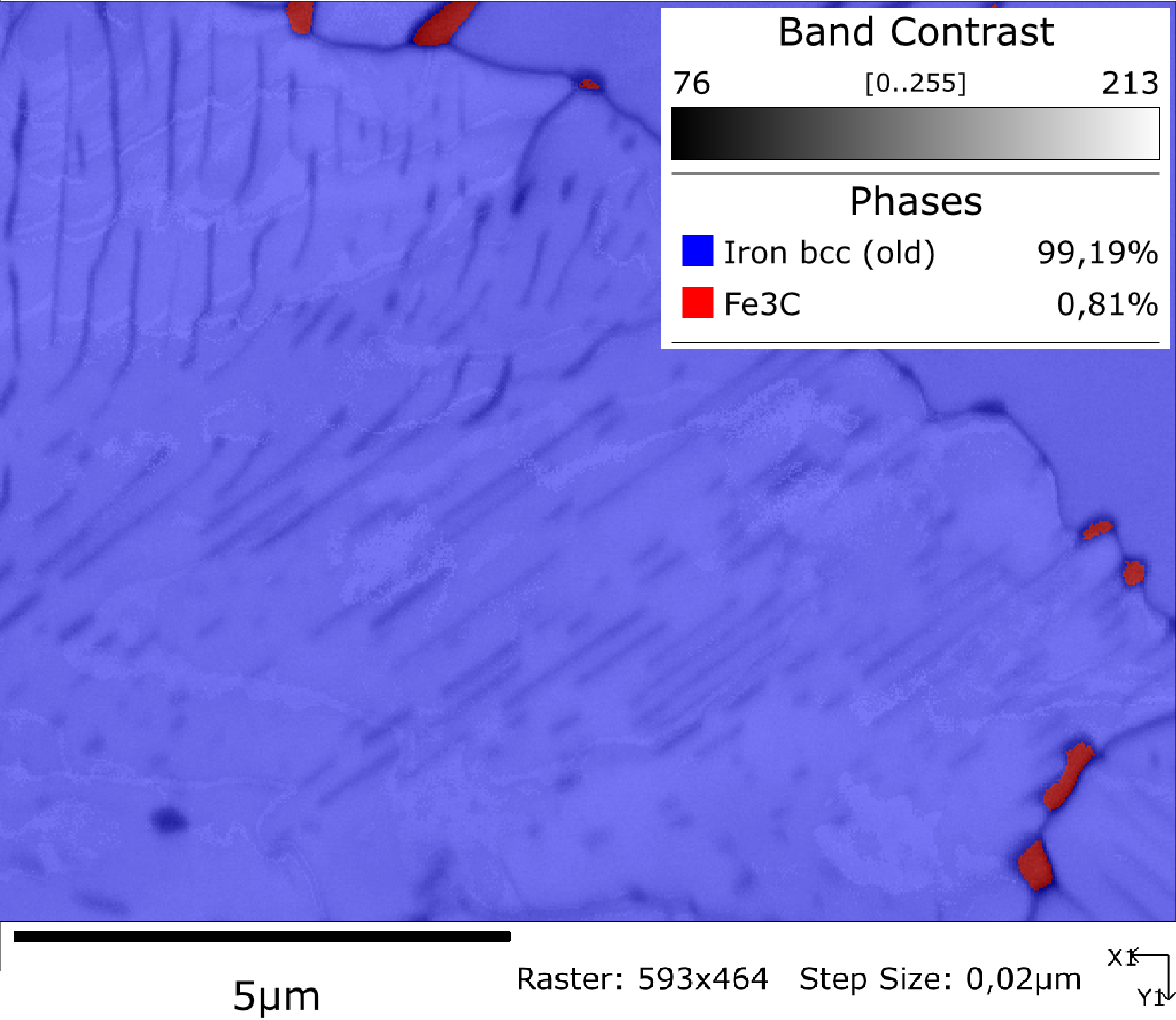}
    \caption{As collected}
    \end{subfigure}    
    \begin{subfigure}{.5\textwidth}
         \includegraphics[clip,width=\textwidth]{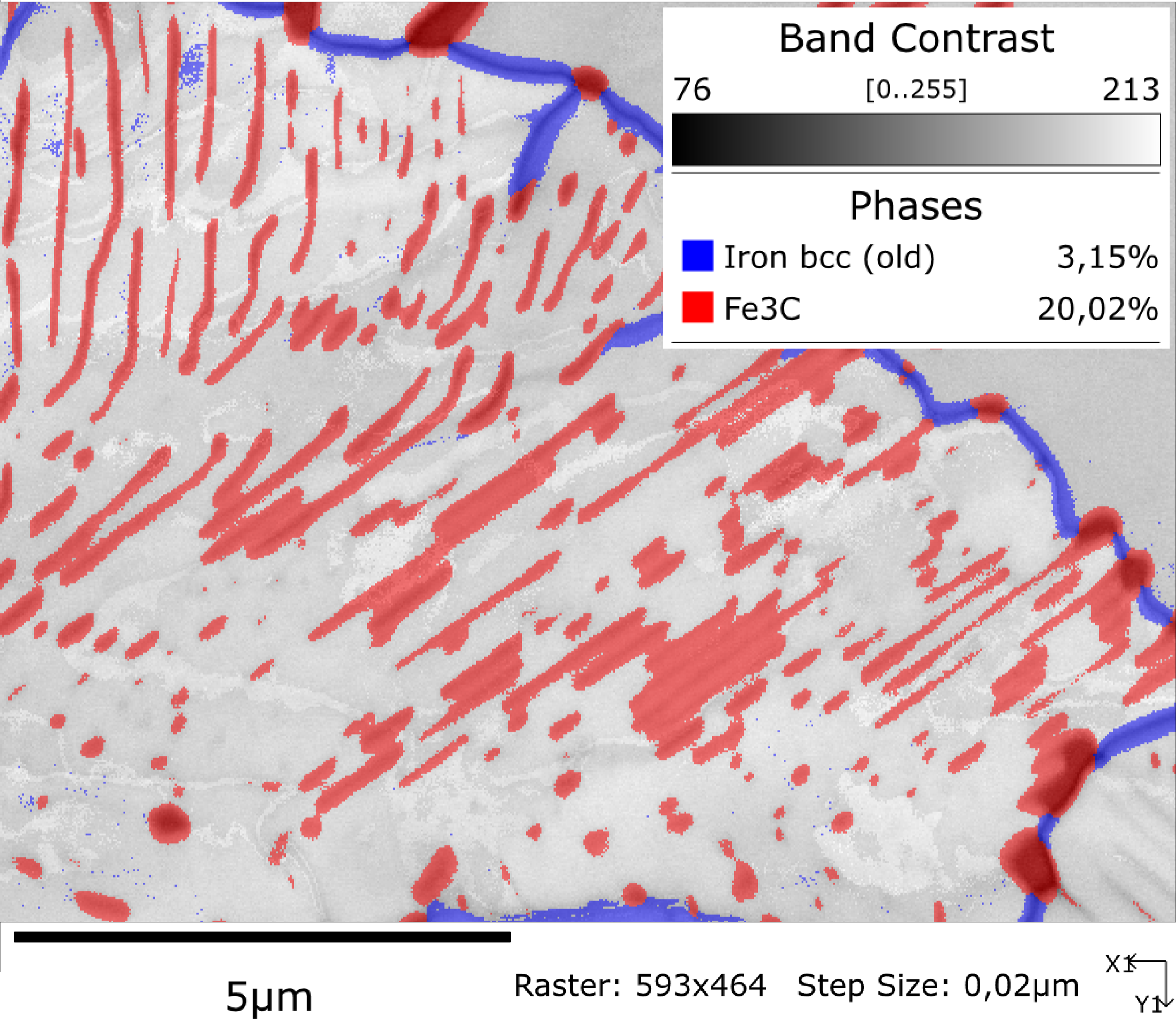}
    \caption{BCC residuals}     
    \end{subfigure} 
\end{center}
    \caption{Ferritic-perlitic steel sample. Phase maps; as received (a) after BCC phase subtraction (b) NCC > 0.175}
    \label{fig:Phase_maps}
\end{figure}

The phase map of ferritic-perlitic steel is shown in Fig.\,\ref{fig:Phase_maps}. Some of the cementite particles on the grain boundaries and the pearlite grain in the left corner were successfully indexed by the Hough-based algorithm. However, the fine pearlite colony in the center of the map remains indexed as BCC ferrite. The underlying band contrast image shows the true nature of the pearlite grain, an eutectoid mixture of ferrite and non-indexed cementite (Fe$_3$C). Band contrast values drops approximately from 190 to 170 from ferrite to cementite lamellae. 
The program was set to subtract only BCC patterns; therefore, after subtracting the strongest signal, the majority of the remaining BCC phase was indexed at the grain boundaries where two BCC orientations overlapped. In the pearlite colony, the major indexed phase was cementite. More than 19\% of the ferrite pixels overlapped with the cementite pattern, and approximately 3\% of the ferrite overlapped with different ferrite orientations at grain boundaries of high and low-angle. The expected value of the volume fraction of cemenetite in pearlite is approximately 12.5\%. The 20\% detected here might be an indication that the lamellae were not perpendicular to the cross section investigated. Pearlite indexed as a mixture of BCC ferrite and Fe$_3$C cementite is a rarity among EBSD analyzes\cite{DURGAPRASAD2017278}. In both examples, samples were prepared by electropolishing or Nital etching, which made the cementite lamellae stick out from the surface and overlap less with the ferrite matrix. The map of the whole colony presented in the present article makes statistically relevant studies on orientation of cementite in pearlite possible. Representative experimental EBSPs from cementite lamellae, together with the corresponding simulations and residual patterns, are shown in the Supplementary Information (Figs. S1 and S2).

\subsection*{Two phase bronze}
\begin{figure}[htb!]
\begin{center}
    \begin{subfigure}{.5\textwidth}
         \includegraphics[clip,width=\textwidth]{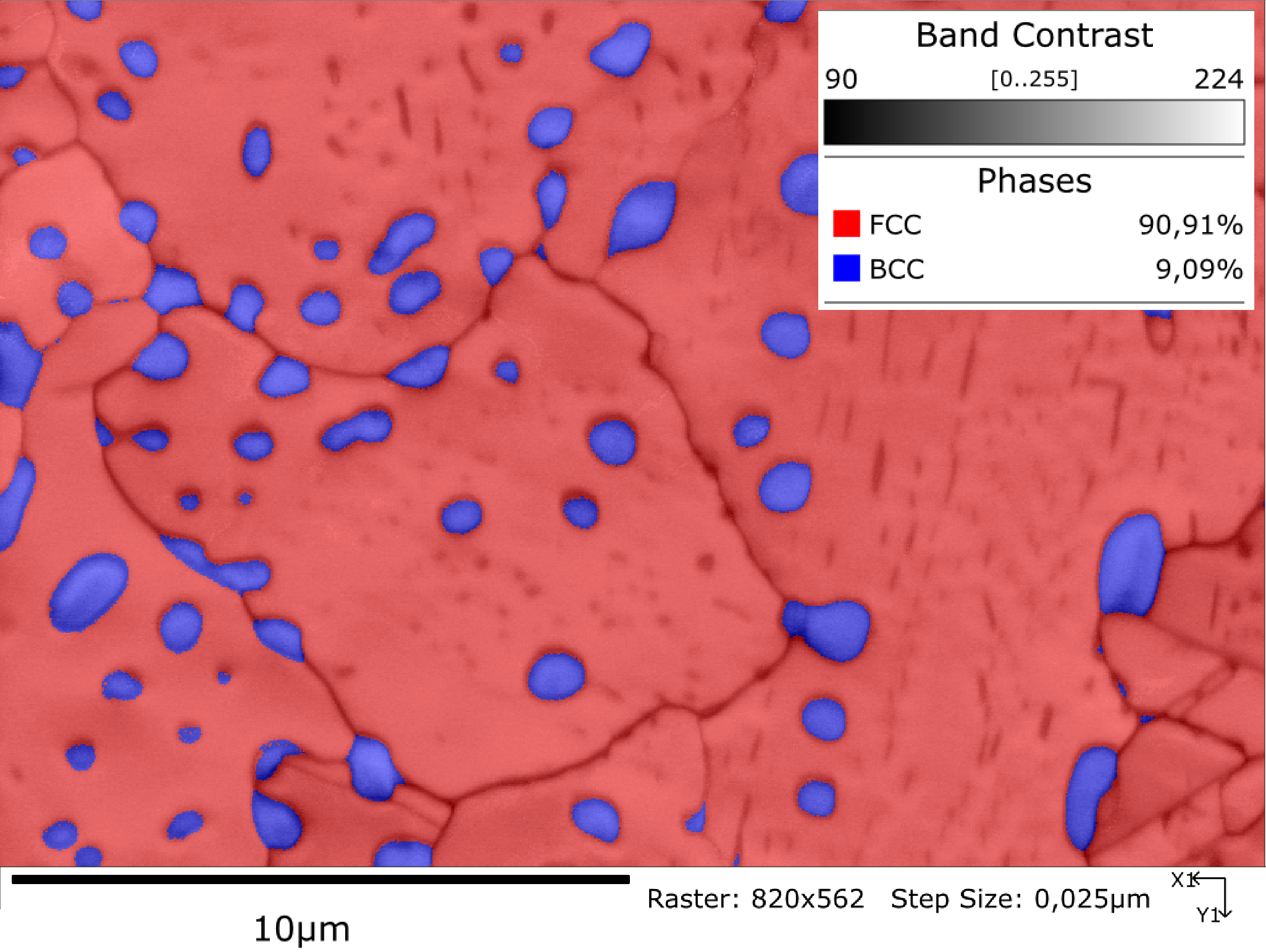}
    \caption{As collected}
    \end{subfigure}    
    \begin{subfigure}{.5\textwidth}
         \includegraphics[clip,width=\textwidth]{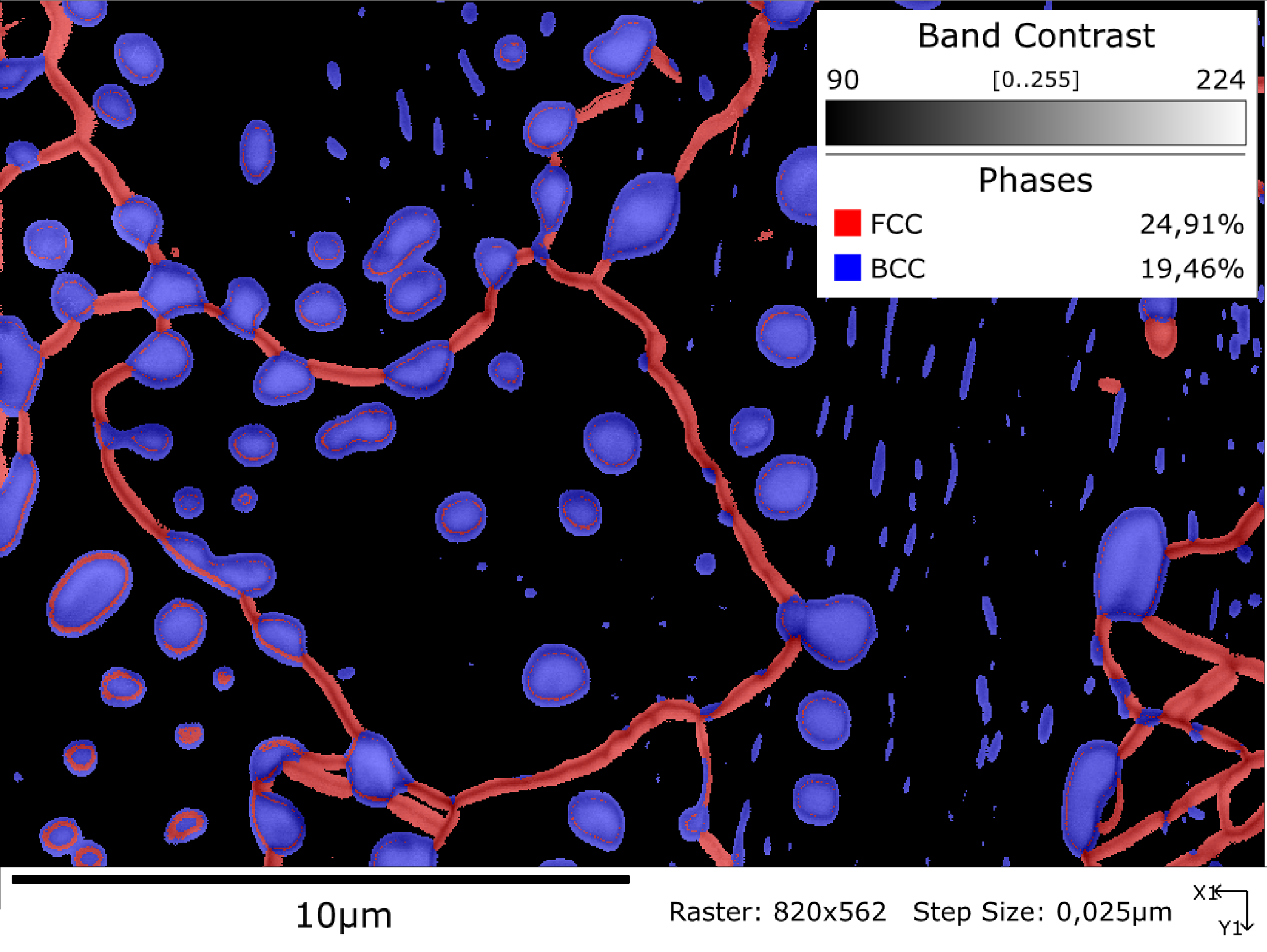}
    \caption{FCC residuals}     
    \end{subfigure} 
\end{center}
    \caption{Cu–Al–Ni–Fe two-phase bronze. Phase maps; as received (a) after FCC phase subtraction (b) NCC > 0.2}
    \label{fig:Phase_maps_Bronze}
\end{figure}

The two phase bronze has a solid FCC copper-based solution; the microstructure also contains a blocky spheroidal beta BCC phase. As can be seen in the Band Contrast overlaid phase map in Fig.\ref{fig:Phase_maps_Bronze}a), there are minor secondary phases that are indicated only by band contrast drops. Their morphology, depending on orientation to the sample surface, is spheroidal and lamellar. After subtraction and DTM these were recognized as beta BCC precipitates (see Fig. {\ref{fig:Phase_maps_Bronze}}b)). Additional EBSPs illustrating the separation of FCC copper and secondary BCC $\beta$-phase contributions are provided in the Supplementary Information (Figs. S5 and S6).

\subsection*{Dual-phase steel}
\begin{figure}[htb!]
\begin{center}
    \begin{subfigure}{.5\textwidth}
         \includegraphics[clip,width=\textwidth]{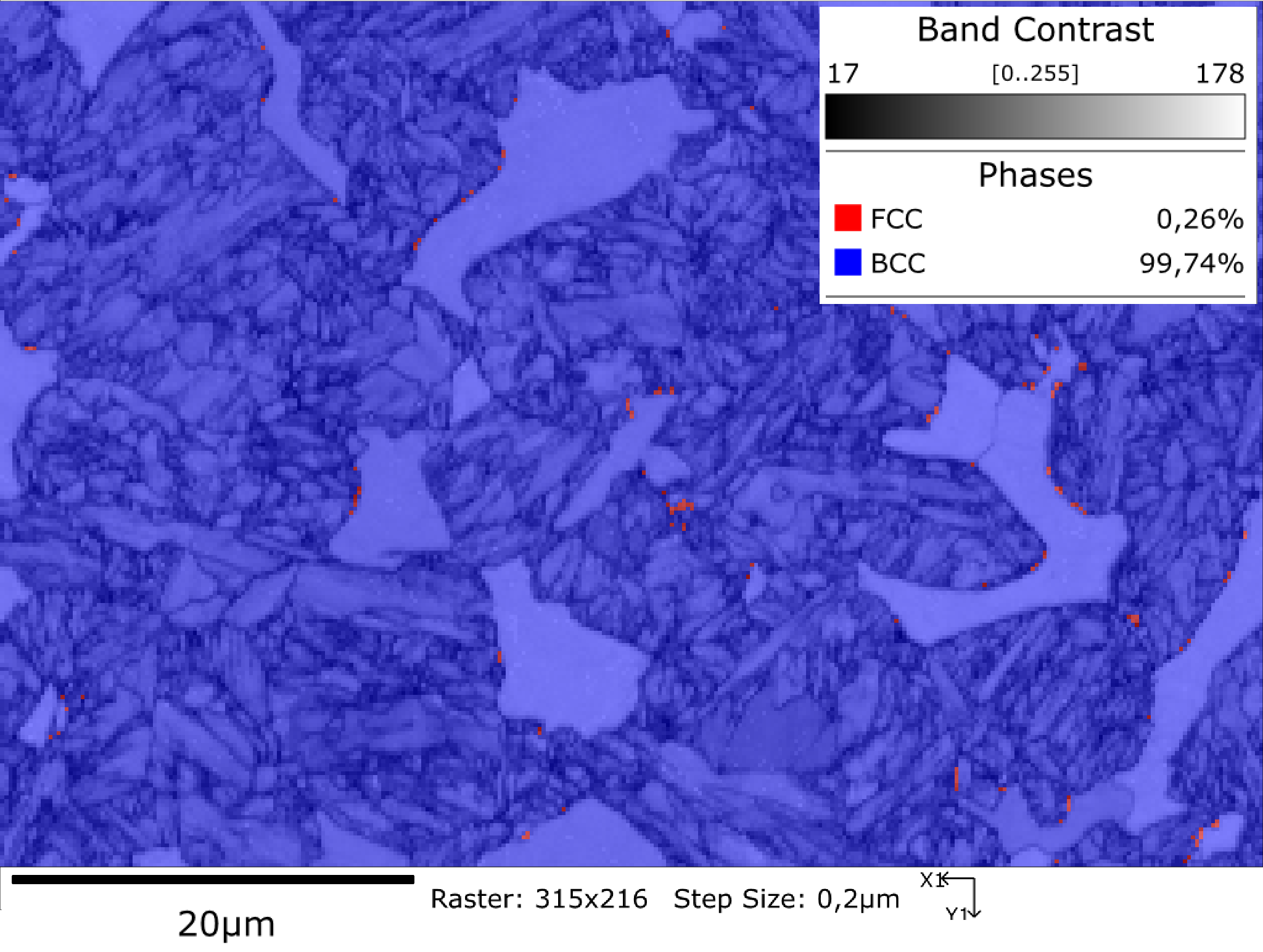}
    \caption{As collected}
    \end{subfigure}    
    \begin{subfigure}{.5\textwidth}
         \includegraphics[clip,width=\textwidth]{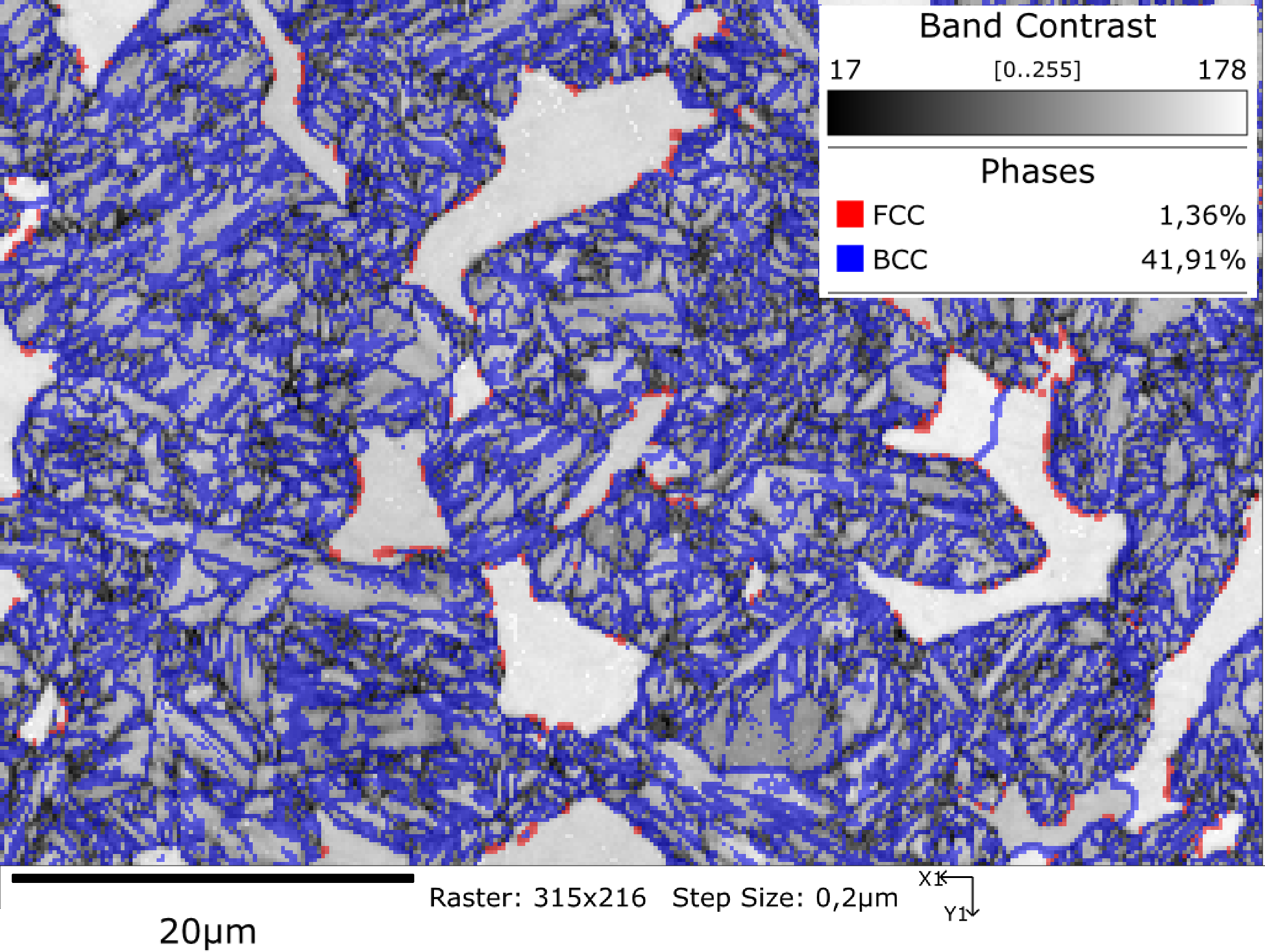}
    \caption{BCC residuals}     
    \end{subfigure} 
\end{center}
    \caption{Dual-phase steel microstructure. Phase maps; as received (a) after BCC phase subtraction (b) NCC > 0.2}
    \label{fig:Phase_maps_DP}
\end{figure}

The sample contains BCC ferrite islands surrounded by BCC/BCT martensite. BCC ferrite can be recognized by higher values of the Band Contrast. In this case, after indexing small quantities of retained austenite of FCC were present \ref{fig:Phase_maps_DP}a). Such scattered pixels and low quantities (0.26\%) may be interpreted as misindexing and disregarded during interpretation. Therefore, the residuals indexing was applied. On the residuals map, the amount of austenite increased by 1.1\% and confirmed the presence of retained austenite. What is interesting is that retained austenite is mainly located at the ferrite-martensite boundaries. Representative experimental and residual EBSPs confirming the presence of retained austenite are shown in the Supplementary Information (Figs. S9 and S10).

\subsection*{Annealed H11 steel}
\begin{figure}[htb!]
\begin{center}
    \begin{subfigure}{.5\textwidth}
         \includegraphics[clip,width=\textwidth]{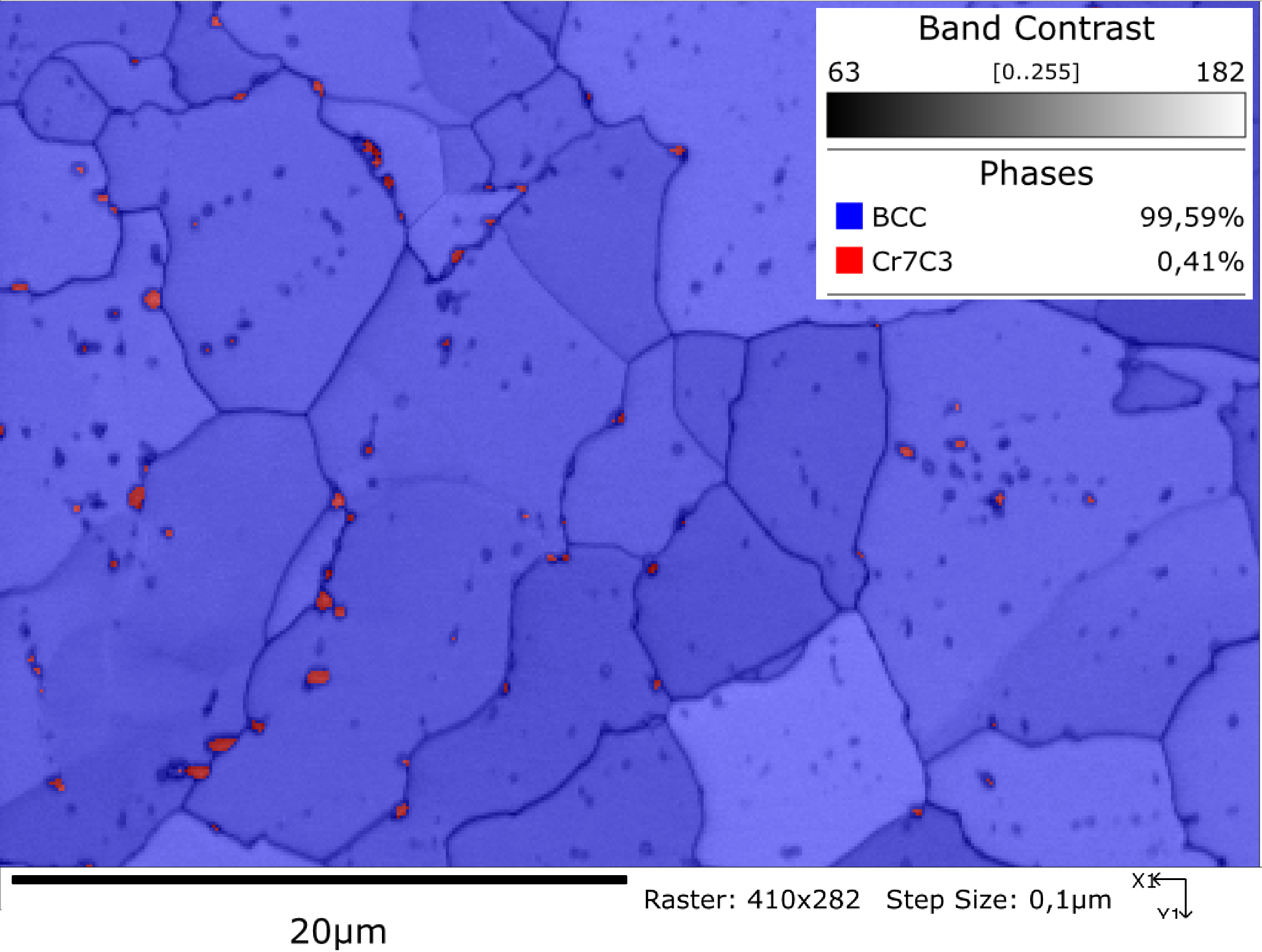}
    \caption{As collected}
    \end{subfigure}    
    \begin{subfigure}{.5\textwidth}
         \includegraphics[clip,width=\textwidth]{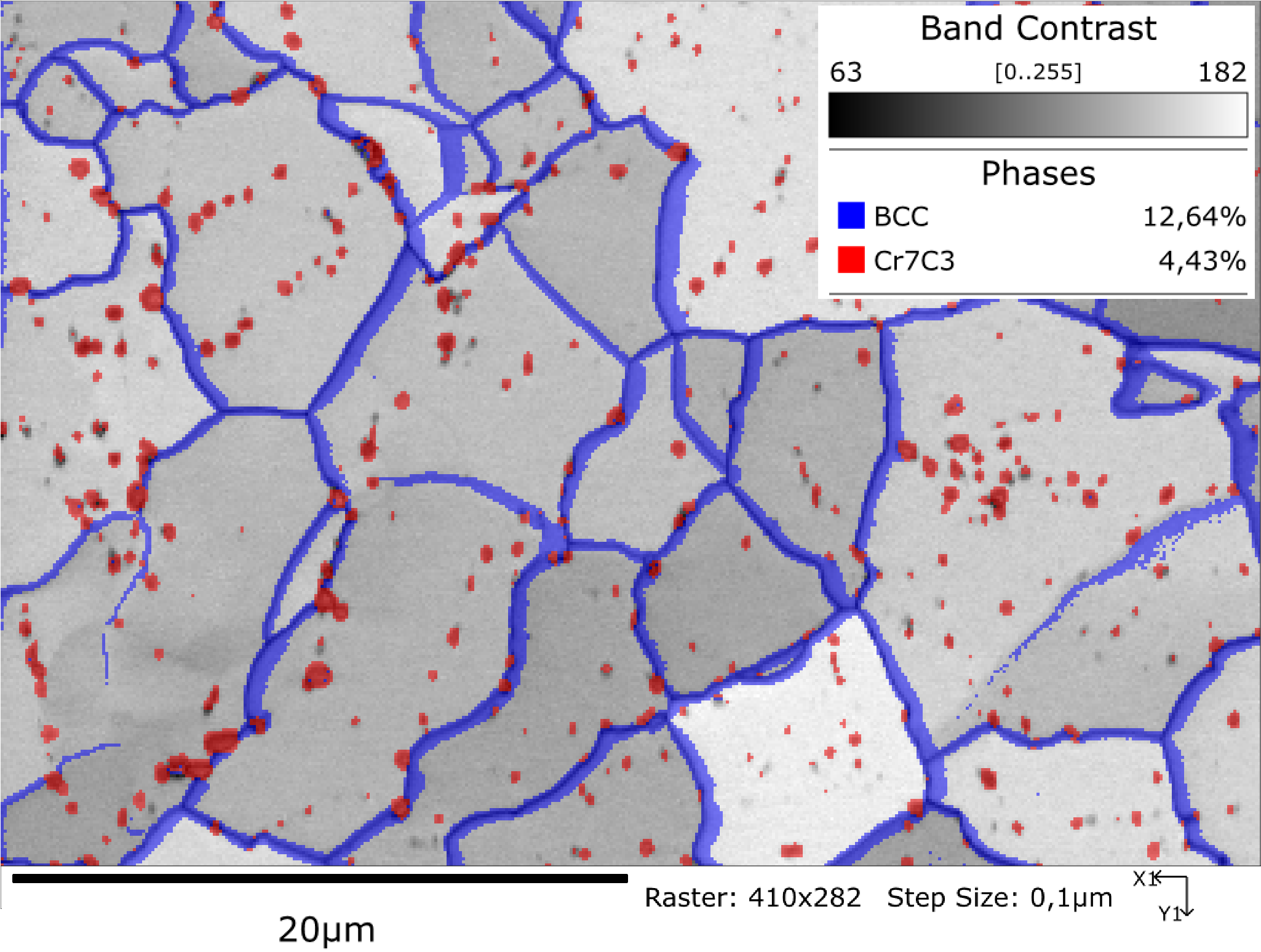}
    \caption{BCC residuals}     
    \end{subfigure} 
\end{center}
    \caption{AISI H11 tool steel. Phase maps; as received (a) after BCC phase subtraction (b) NCC > 0.2}
    \label{fig:Phase_maps_CR7C3}
\end{figure}

In the tool steel microstructure the fraction of the initial identified Cr$_7$C$_3$ carbides was 0.41\% (see Fig.\ref{fig:Phase_maps_CR7C3})a, while assuming that steel consists only of Cr$_7$C$_3$ carbides, the volume fraction of carbides is expected to be in the range of 4.25 to 5.27\%, which can be slightly overestimated due to the presence of small amounts of V and Mo carbides. After subtracting the BCC phase patterns and indexing the residuals by DTM, the identified volume fraction of Cr$_7$C$_3$ carbides was 4.43\% which fits better to the estimated values of equilibrium conditions presented above (see Fig.\ref{fig:Phase_maps_CR7C3})b). Representative EBSPs acquired from Cr$_7$C$_3$ carbides, together with the corresponding simulations and residual patterns, are provided in the Supplementary Information (Figs. S3 and S4).

\subsection*{Supposedly epitaxial Cu thin film}
\begin{figure}[htb!]
\begin{center}
    \begin{subfigure}{.5\textwidth}
         \includegraphics[clip,width=\textwidth]{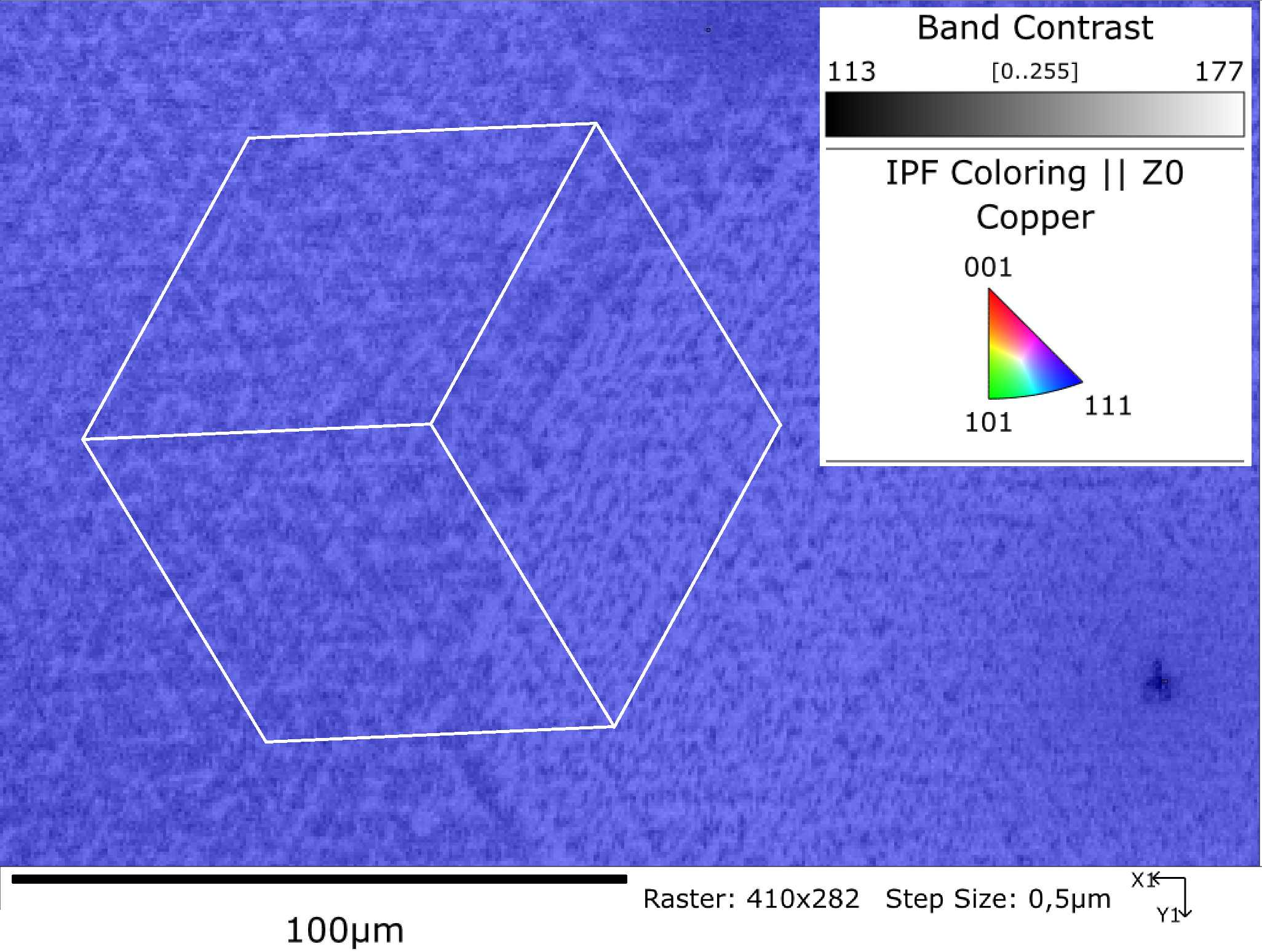}
    \caption{As collected}
    \end{subfigure}    
    \begin{subfigure}{.5\textwidth}
         \includegraphics[clip,width=\textwidth]{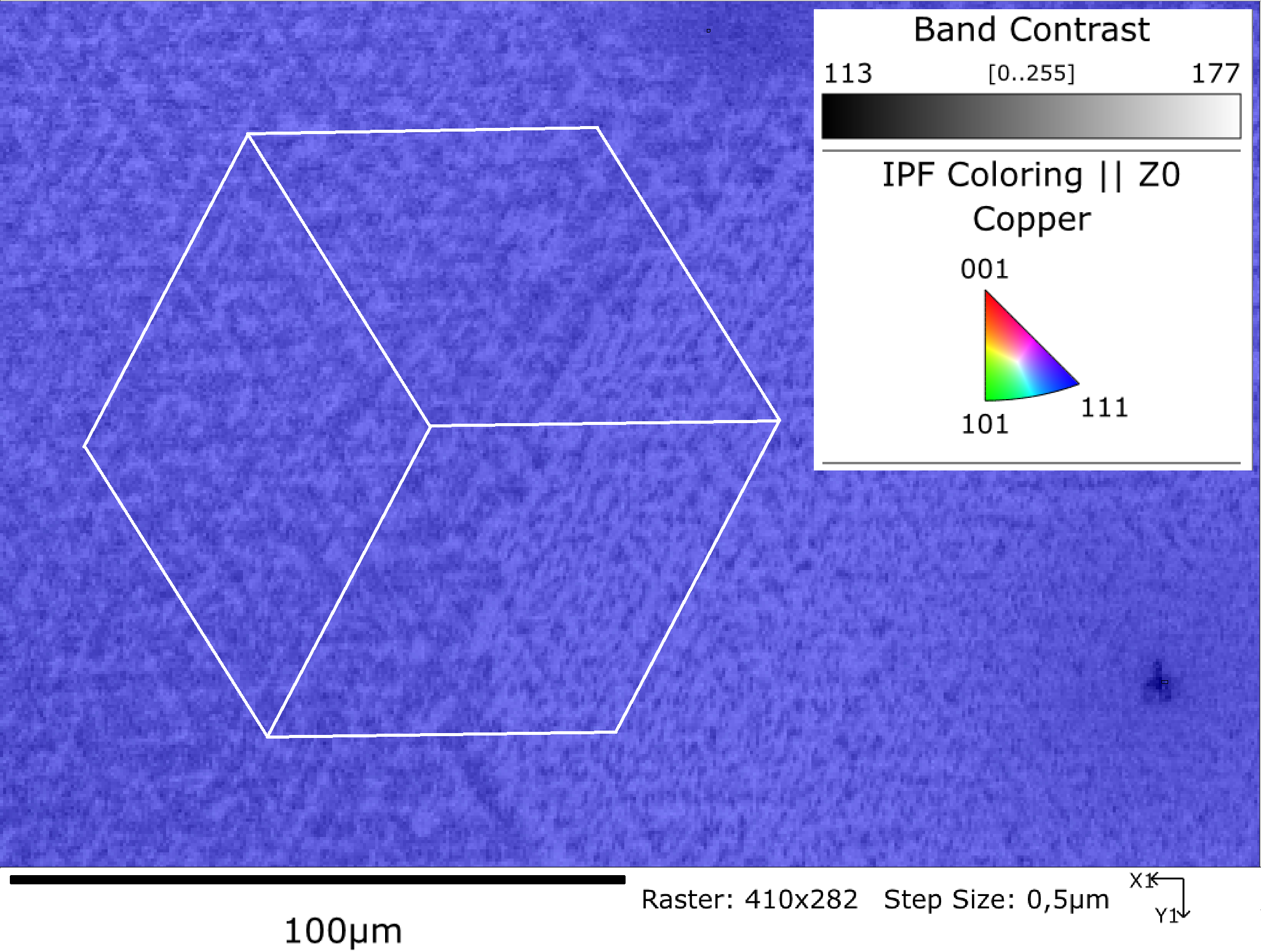}
    \caption{FCC residuals}     
    \end{subfigure} 
\end{center}
    \caption{CuPt thin film on (111) STO substrate. IPF maps; as received (a) after FCC phase subtraction (b) NCC > 0.2}
    \label{fig:Phase_maps_thi_film}
\end{figure}

After classical Hough transform based indexing as well as DTM the sample was identified as single crystalline - i.e. one orientation as seen in the Fig.\ref{fig:Phase_maps_thi_film})a with overlaid crystal frame. However, after subtracting the strongest signal from each map pixel EBSP it turned out that each pattern consisted of two 60\degree{} <111> twin related (see Fig.\ref{fig:Phase_maps_thi_film})b). Representative experimental and residual EBSPs demonstrating the separation of twin-related FCC contributions are shown in the Supplementary Information (Figs. S7 and S8). In other words in every single EBSD map pixel electron matter interaction volume two twin related domains were present making the film far from epitaxial.

\subsection*{Discussion}
The presented methodology overcomes the limitations of previous approaches such as Lenthe et al. \cite{Lenthe2020} and Brodu et al. \cite{Brodu2022}. In \cite{Lenthe2020}, the successful resolution of the overlapping patterns was possible because an orientation relationship between the two phases analyzed was known and could be exploited. However, in many real microstructures, such orientation relationships are absent or highly variable. For example, the ferritic steel microstructure with Cr$_7$C$_3$ carbides analyzed here represents a case where no specific orientation relationship is expected between matrix and carbides, yet the proposed residual-based method still provides reliable results. From the computational point of view, attempting to account for all possible orientation variants using dictionary indexing or spherical indexing would require repeating the analysis many times (up to 24 iterations for the case of Kurdjumov–Sachs relationship), making such an approach computationally demanding and impractical for routine analysis. Normalized cross-correlation (R) maps for experimental and residual EBSPs of selected material systems are provided in the Supplementary Information (Figs. S11–S20).

In contrast, the method of Brodu et al. \cite{Brodu2022} does not require a known orientation relationship between overlapping patterns, but it does rely on the availability of at least one non-overlapping experimental pattern with a similar orientation. This clean reference pattern can then be subtracted from the overlapping measurement to reveal the weaker signal. It should be noted that this methodology was later submitted for patent (US20240186105A1). However, its applicability is inherently limited to situations where such non-overlapping patterns exist within the dataset.

Our approach is fundamentally different. Instead of requiring an additional experimental reference, it employs dynamically simulated Kikuchi patterns as the subtraction reference. The subtraction is guided by physics-based forward modeling and optimized mathematically using normalized cross-correlation scaling, Gaussian blurring, and gain correction. This strategy makes the method applicable even in cases where all measured EBSPs are overlapping. The copper thin film example illustrates this advantage clearly: every measured pattern in this sample contained signals from two 60\degree{} around <111> twin-related domains, which means that no purely experimental subtraction method could succeed. In contrast, our simulation-driven residual analysis successfully revealed twin-related domains, demonstrating that experiment–experiment subtraction strategies are not viable in such a case.

This highlights a fundamental distinction: by relying on simulations rather than neighboring experimental patterns, the proposed approach does not assume the existence of a dominant clean signal in the dataset, nor does it require a predefined orientation relationship. Instead, it provides a general and transferable framework that can be applied pixel by pixel in a wide range of microstructures.

This approach also suggests a new way of thinking about EBSD data. By performing iterative subtraction, one can in principle generate residuals of the first, second, third, or even higher order. In complex microstructural configurations, such as particles located at grain triple junctions, this could allow the separation of multiple crystallographic contributions within a single diffraction pattern. Such developments may transform EBSD maps into a kind of hyperspectral dataset, where each pixel contains not only a dominant orientation but also a hierarchy of weaker residual signals. This perspective would require new software tools for analysis, since for example crystallographic orientation relationships could then be determined inside a single pixel rather than only between neighboring pixels. Users could further define quantitative parameters based on the relationships between different residual levels, opening a path toward a more detailed description of local microstructure than has been possible with conventional EBSD.

\section{Summary}
The methodology presented in this work enables a reliable separation of overlapping EBSD patterns through experiment–simulation residual analysis. By combining dynamic template matching with optimized subtraction using normalized cross-correlation scaling, Gaussian blurring, and gain correction, the approach enhances the detection of weak secondary signals and minor phases that are otherwise obscured by dominant diffraction patterns. The method was successfully demonstrated on steels, bronze, and thin film, providing improved identification of cementite lamellae, retained austenite, carbides, and twin-related domains. These results confirm that simulation-driven residual analysis can significantly extend the effective spatial resolution of EBSD and open new possibilities for the microstructural characterization of complex materials.


\FloatBarrier
\bigskip

\section*{Acknowledgments}
This work was supported by the Polish National Science Centre (NCN), grant no. 2020/37/B/ST5/03669.

The research results presented in this paper have been developed with the use of equipment financed from the funds of the "Excellence Initiative - Research University" program at the AGH University of Krakow. The research project is partly supported by the program "Excellence initiative – research university” for the AGH University. The authors thank Dr Piotr Salwa, Dr Piotr Jabłoński (AGH University of Krakow) and Dr Meysam Haghshenas (University of Toledo) for providing the dual-phase steel, CuPt thin film and two-phase bronze sample, respectively. 

\section*{Data availability}
The raw data and the code required to reproduce the above findings are available for download from Zenodo \href{https://doi.org/10.5281/zenodo.16746462}{DOI: 10.5281/zenodo.10808974} and \href{https://doi.org/10.5281/zenodo.17079414}{DOI:10.5281/zenodo.17079414}

\section*{Declaration of competing interest}  
The authors declare that they have no known competing financial interests or personal relationships that could have appeared to influence the work reported in this paper.



\bibliographystyle{elsarticle-num}
\bibliography{references}

\section*{Authors contribution}
G.C Conceptualization; Investigation; Software; Project administration; Visualization; Writing - original draft; Writing - review \& editing\newline
A.W. Software; Validation; Visualization; Writing - review \& editing; Conceptualization; Funding acquisition;\newline
T.T. Supervision; Validation\newline
W.B. Investigation;\newline
P.B. Funding acquisition; Supervision\newline

\section*{Appendix}

\subsection*{Normalized cross-correlation coefficients and residual patterns}

In this appendix, we demonstrate a result for subtraction of a pattern simulation from the experiment.  
When comparing experimental and simulated Kikuchi patterns, we can ask which scaled simulation provides the best least-squares approximation to the signal under the condition that the normalized cross-correlation coefficient $r$ between experimental and simulated patterns has been maximized.
We consider an idealized situation where all pixels of the measured Kikuchi patterns \(T\) and the simulation \(S\) contribute with equal signal-to-noise ratio.

The two signals \(T\) and \(S\) each are normalized to have mean \(0.0\) and standard deviation \(1.0\). The normalized cross-correlation coefficient between \(T\) and \(S\) then is given by
\[
r = \frac{E[T \cdot S]}{\sigma_T \sigma_S} = E[T \cdot S],
\]
since \(\sigma_T = \sigma_S = 1\). 

We wish to find the scaling factor \(a\) and offset \(m\) such that the expected squared error
\[
f(a, m) = E\left[(T - a\,S + m)^2\right].
\]
is minimized.
This can be approached by expanding the square:
\[
\begin{aligned}
f(a, m) &= E\left[(T - aS + m)^2\right] \\
&= E\Bigl[T^2 - 2a\,T\,S + a^2 S^2 + 2m\,T - 2a\,m S + m^2\Bigr].
\end{aligned}
\]

Using the linearity of the expectation, we have:
\begin{align*}
f(a, m) &= E[T^2] - 2a\,E[T\,S] + a^2E[S^2]  \\ 
    &\quad + 2m\,E[T] - 2a\,m\,E[S] + m^2
\end{align*}

We can substitute the known statistics, given the properties of \(T\) and \(S\):
\begin{itemize}
    \item \(E[T] = 0\), (mean is 0)
    \item \(E[S] = 0\),
    \item \(E[T^2] = 1\) (standard deviation is 1),
    \item \(E[S^2] = 1\),
    \item \(E[T\,S] = r\) (by the definition of the normalized cross-correlation coefficient),
\end{itemize}
The expression for the squared error simplifies to:
\[
f(a, m) = 1 - 2a\,r + a^2 + m^2.
\]

To find the optimal \(m\), we take the partial derivative of \(f(a, m)\) with respect to \(m\):
\[
\frac{\partial f}{\partial m} = 2m.
\]
Setting the derivative equal to zero:
\[
2m = 0 \quad \Longrightarrow \quad m = 0.
\]

Next, we take the partial derivative of \(f(a, m)\) with respect to \(a\):
\[
\frac{\partial f}{\partial a} = -2r + 2a.
\]
Setting the derivative equal to zero:
\[
-2r + 2a = 0 \quad \Longrightarrow \quad a = r.
\]

The above derivation shows that the parameters that minimize the sum of squared differences are:
\[
\boxed{a = r \quad \text{and} \quad m = 0.}
\]

Thus, scaling the simulation \(S\) by the cross-correlation coefficient \(r\) itself, with no offset, provides the best least-squares approximation to the experimental signal \(T\).
This result is in line with general expectations. 
For example, if the simulation fits insufficiently, with only a small $r$, we should also subtract less of the ideal simulated signal from the actual experimental data. 
In the limit of $r=0$, the complete experimental measurement would remain, illustrating that the specific simulation that was used for the fit does not explain anything about the measured pattern ($r=0$).
Also, as both signals are centered around zero, one would not see the necessity of an offset.

\clearpage
\appendix
\section*{Supplementary Information}



\section*{Supplementary material}

**Supplementary Information**

Resolving Overlapping EBSD Patterns by Experiment - Simulation Residuals Analysis

Grzegorz Cios, Aimo Winkelmann, Tomasz Tokarski, Wiktor Bednarczyk, Piotr Bała

**Supplementary Information**

This Supplementary Information provides additional experimental Electron BackScatter Patterns (EBSPs), corresponding best-fit simulations, residual patterns, and normalized cross-correlation (R) maps that support the results and discussion presented in the main manuscript. The figures illustrate representative examples for all investigated material systems and demonstrate the robustness of the simulation-driven residual analysis across different microstructures.

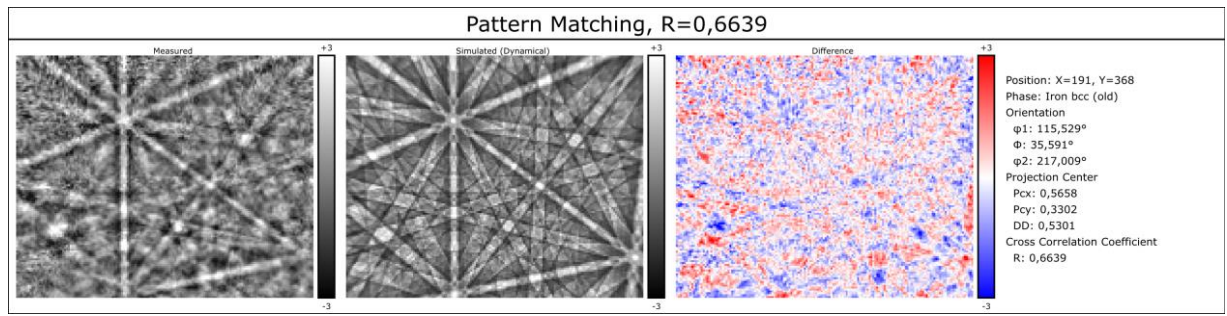

*Figure S1 Experimental EBSP acquired from a cementite lamella in ferritic–pearlitic steel, corresponding best-fit BCC ferrite simulation, and the resulting difference image.*

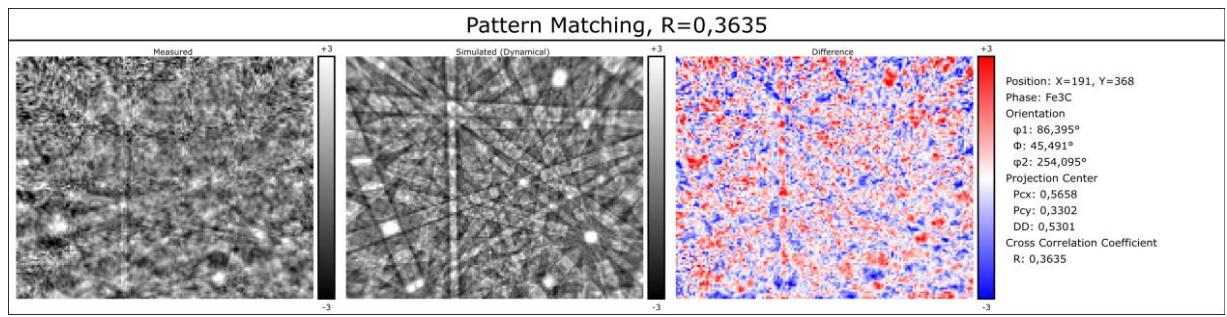

*Figure S2 Residual EBSP obtained after subtraction of the best-fit BCC ferrite simulation shown in Figure S1, together with the best-fit $Fe_3C$ simulation and the corresponding difference image.*

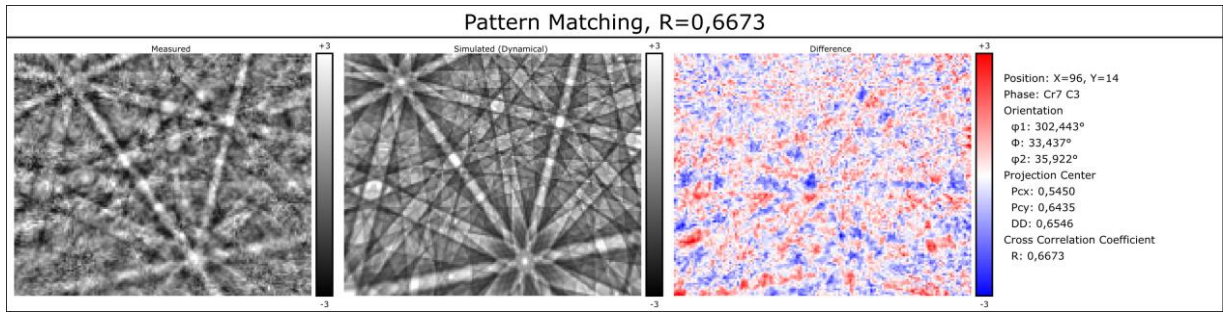

*Figure S3 Experimental EBSP acquired from a $Cr_7C_3$ carbide particle in annealed H11 steel, corresponding best-fit BCC ferrite simulation, and the resulting difference image.*

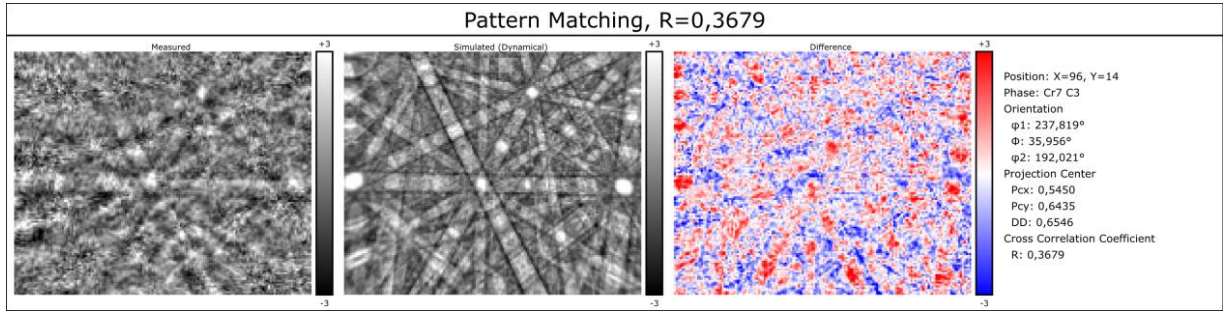

*Figure S4 Residual EBSP after subtraction of the BCC ferrite contribution shown in Figure S3, together with the best-fit $Cr_7C_3$ simulation and the corresponding difference image.*

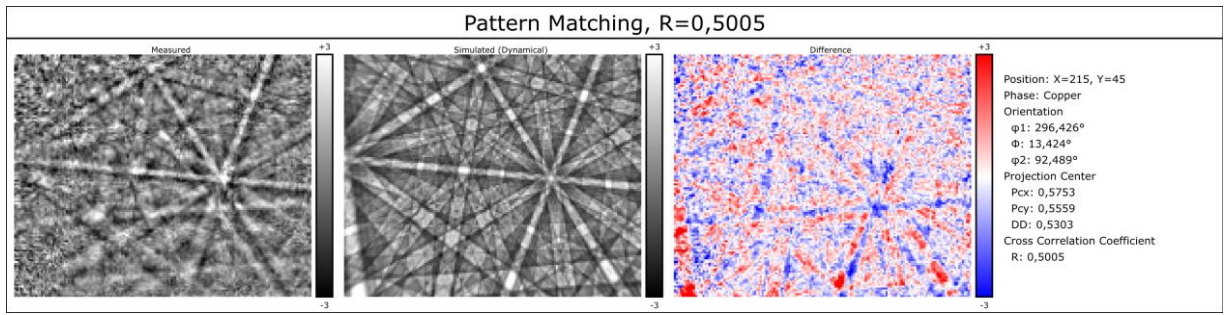

*Figure S5 Experimental EBSP acquired from a secondary BCC β-phase precipitate in the Cu–Al–Ni–Fe bronze, corresponding best-fit FCC copper simulation, and the resulting difference image.*

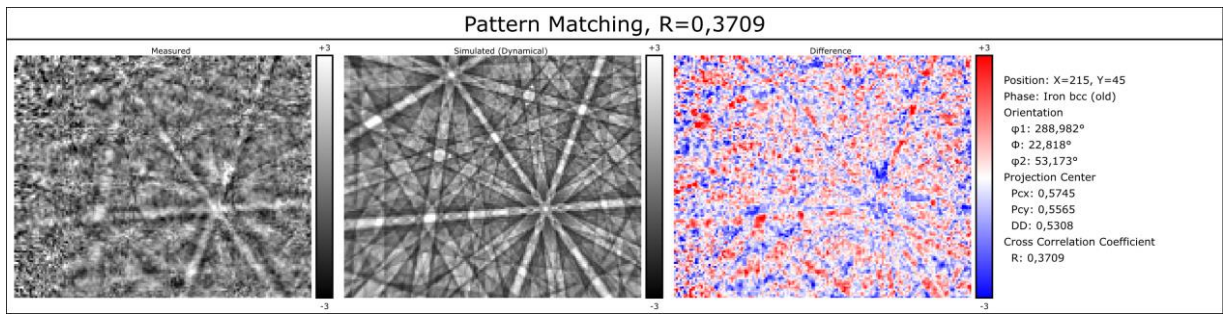

*Figure S6 Residual EBSP obtained after subtraction of the FCC copper contribution shown in Figure S5, together with the best-fit BCC β-phase simulation and the corresponding difference image.*

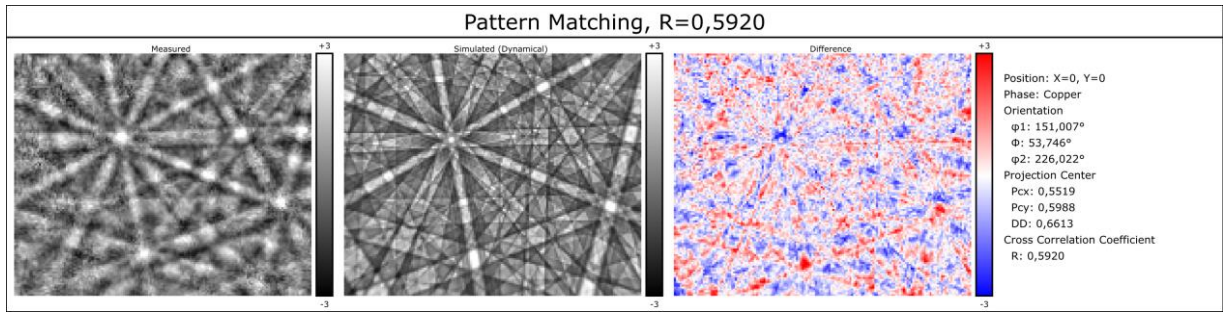

*Figure S7 Experimental EBSP acquired from the Cu thin film deposited on (111) LSMO, corresponding best-fit FCC copper simulation, and the resulting difference image.*

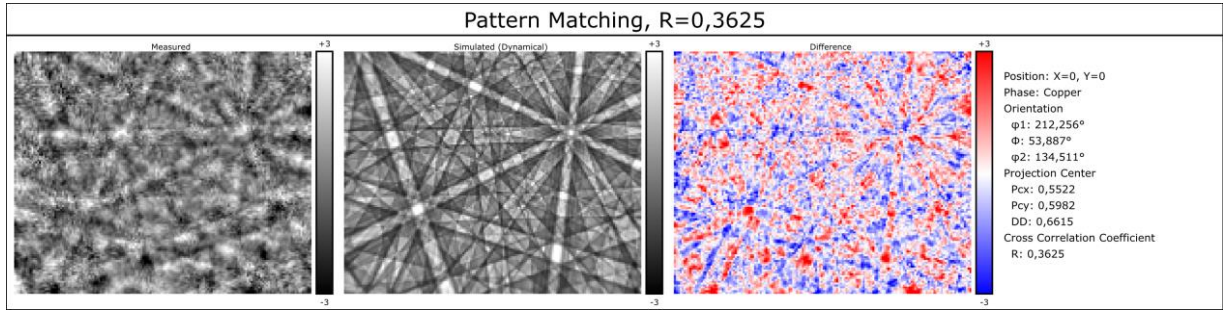

*Figure S8 Residual EBSP obtained after subtraction of the dominant FCC contribution shown in Figure S7, together with the best-fit twin-related FCC simulation and the corresponding difference image.*

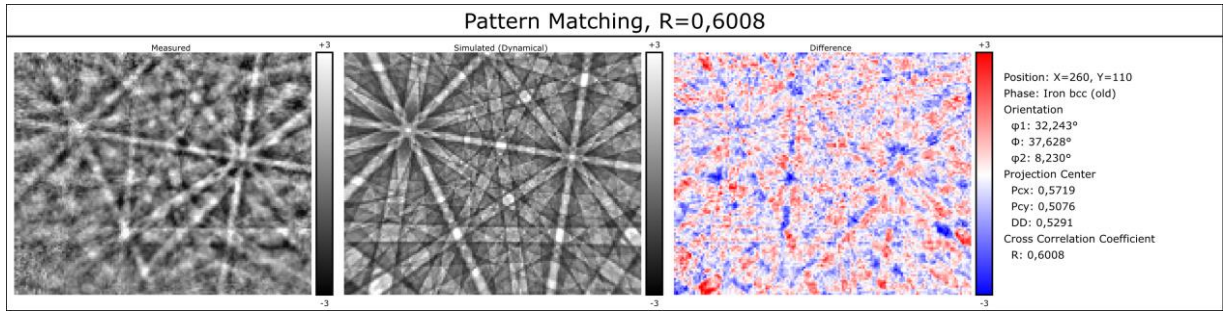

*Figure S9 Experimental EBSP acquired from retained austenite in the dual-phase steel, corresponding best-fit BCC ferrite simulation, and the resulting difference image.*

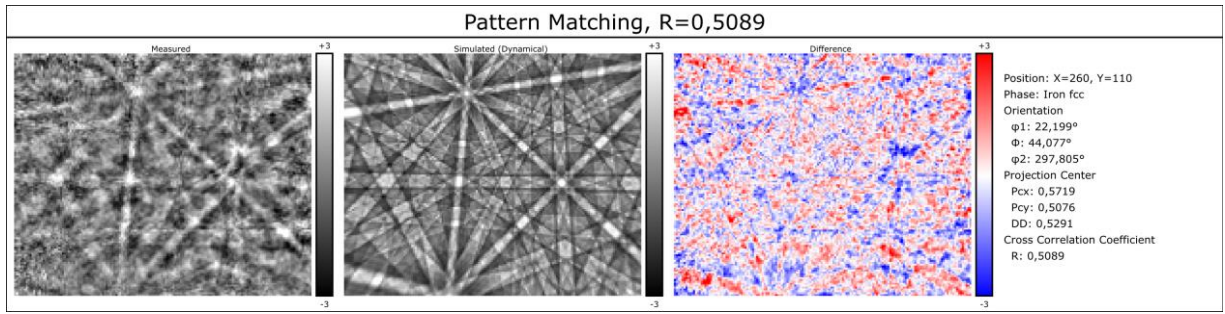

*Figure S10 Residual EBSP obtained after subtraction of the BCC ferrite contribution shown in Figure S9, together with the best-fit FCC austenite simulation and the corresponding difference image.*

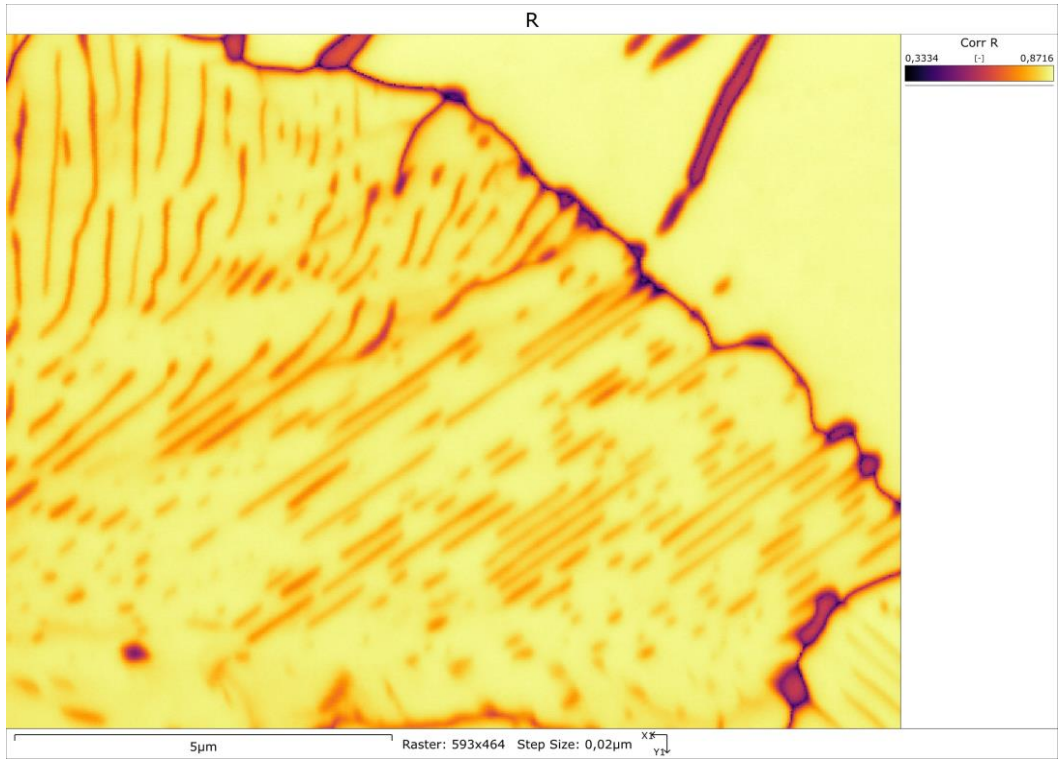

*Figure S11 Normalized cross-correlation (R) maps calculated for experimental EBSPs acquired from ferritic–pearlitic steel prior to residual analysis.*

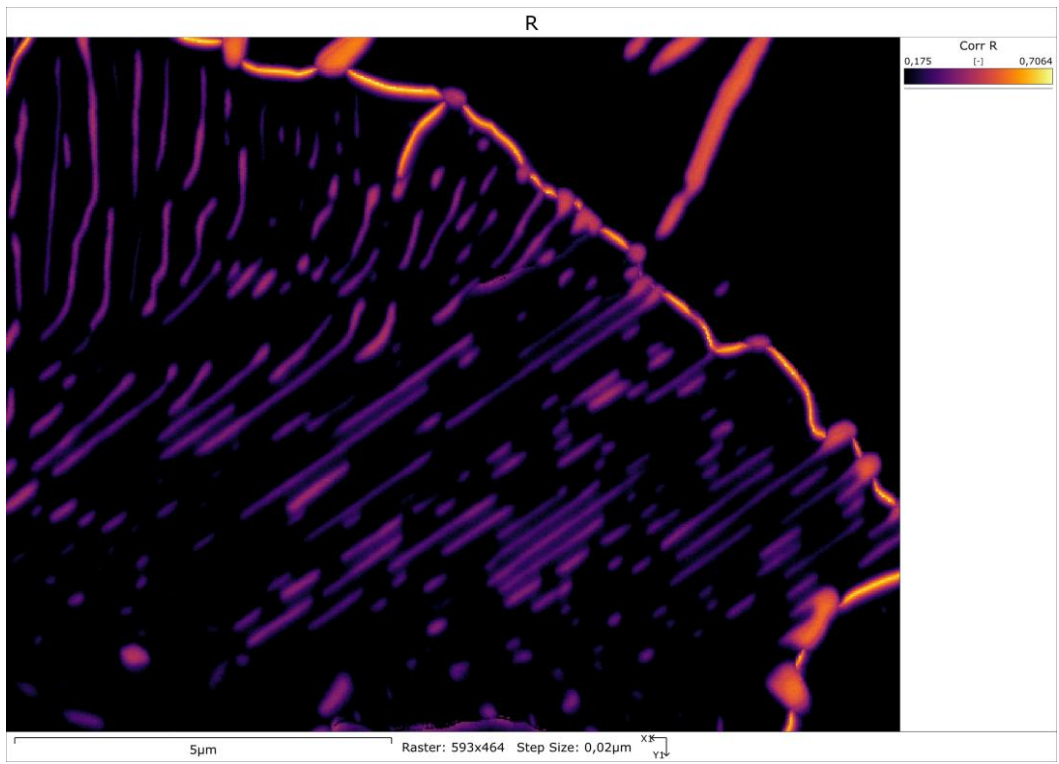

*Figure S12 Normalized cross-correlation (R) maps calculated for residual EBSPs of ferritic–pearlitic steel after subtraction of the dominant BCC ferrite contribution.*

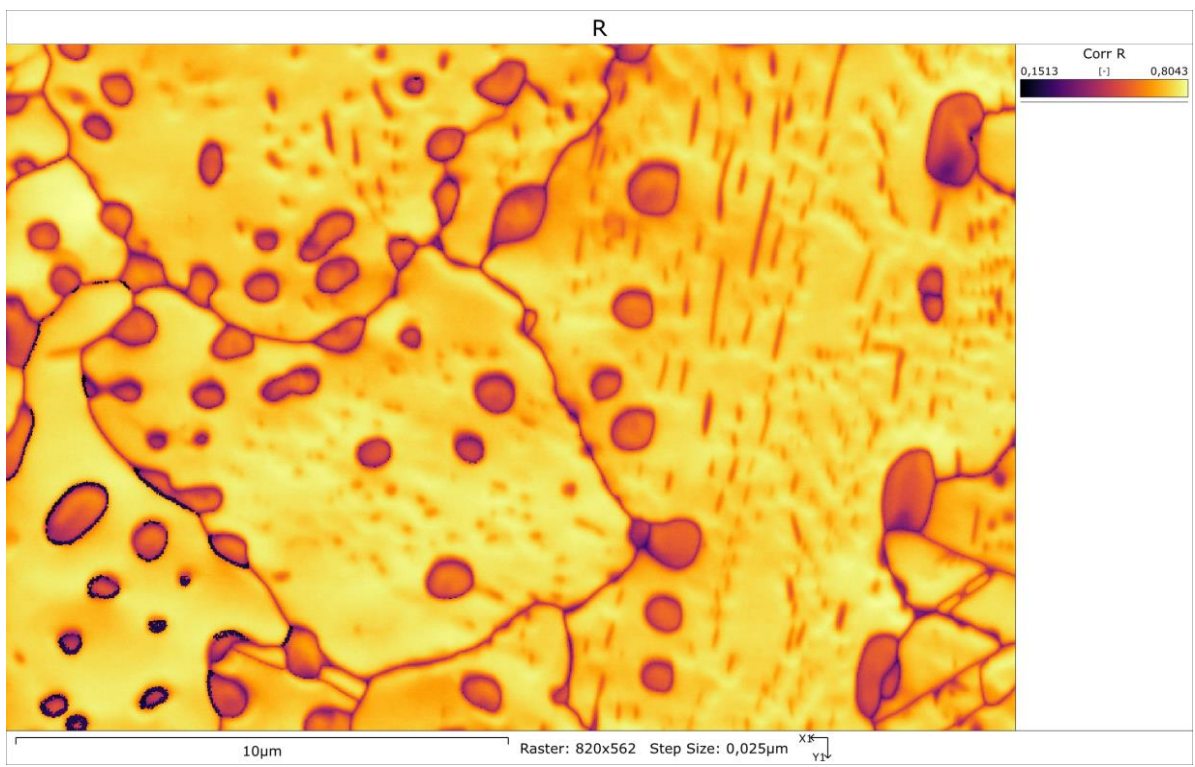

*Figure S13 Normalized cross-correlation (R) maps calculated for experimental EBSPs acquired from the Cu–Al–Ni–Fe two-phase bronze prior to residual analysis.*

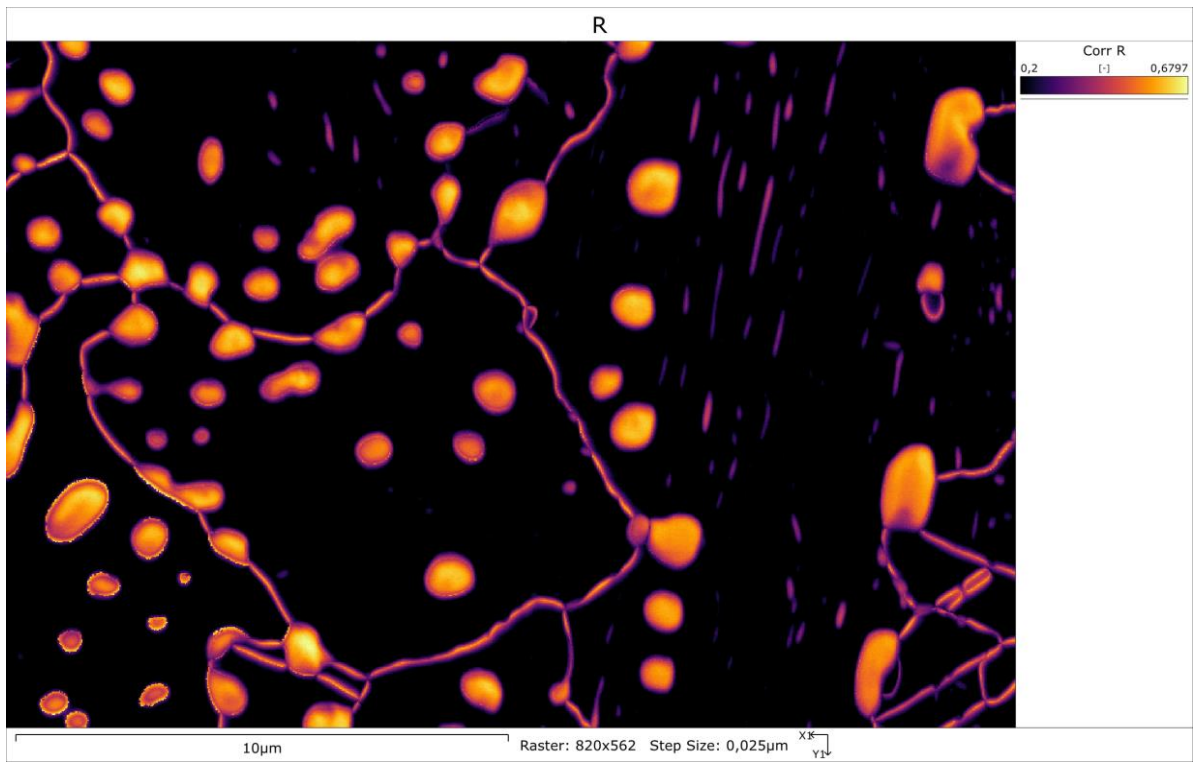

*Figure S14 Normalized cross-correlation (R) maps calculated for residual EBSPs of the Cu–Al–Ni–Fe bronze after subtraction of the dominant FCC copper contribution.*

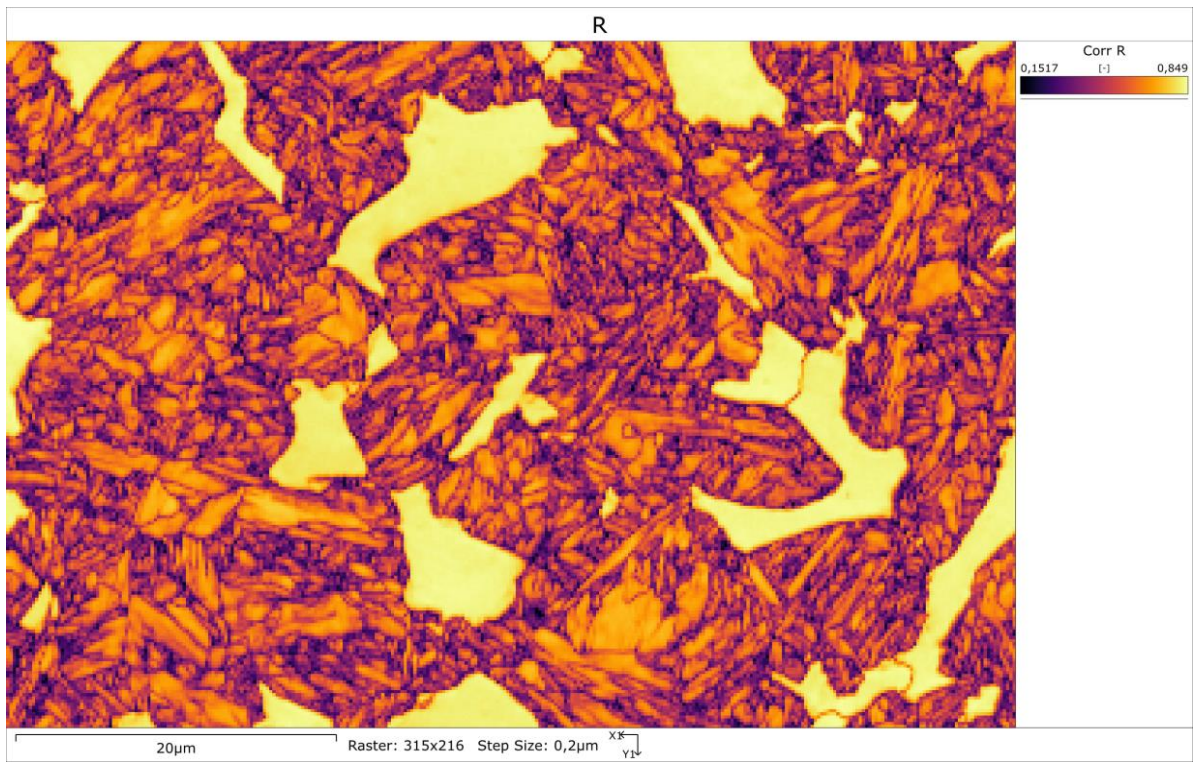

*Figure S15 Normalized cross-correlation (R) maps calculated for experimental EBSPs acquired from the dual-phase steel prior to residual analysis.*

16

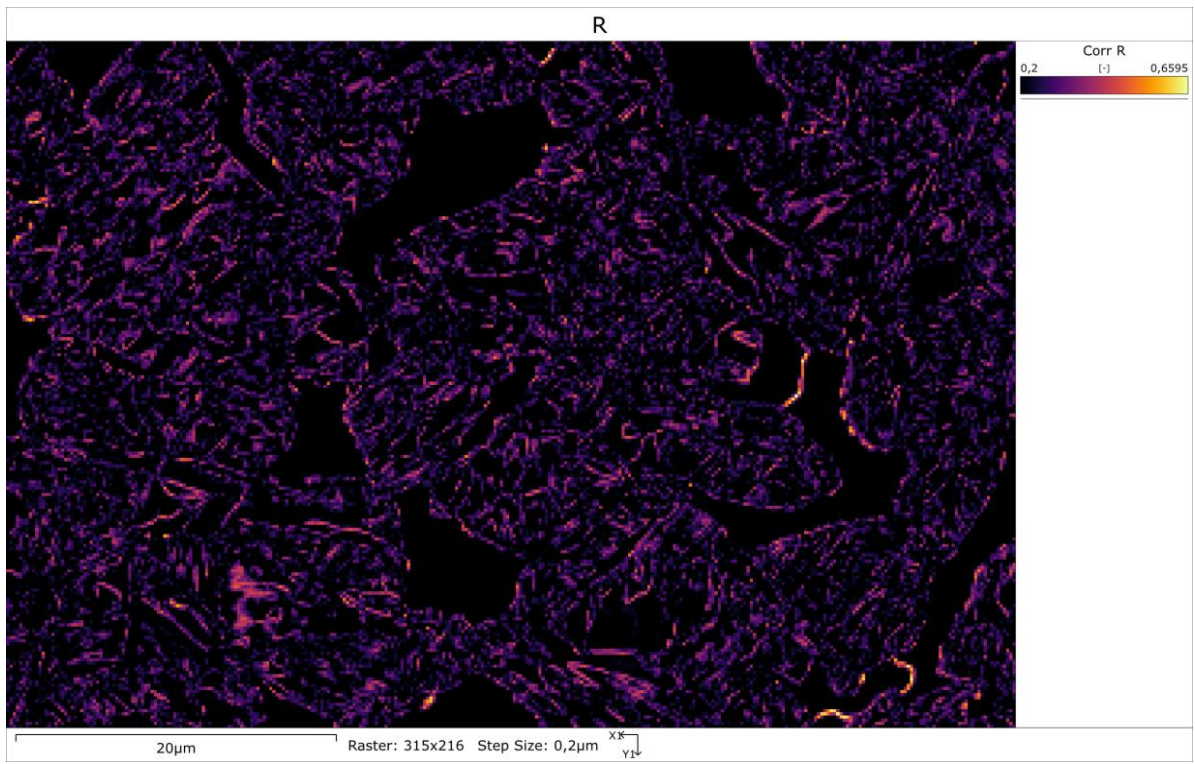

*Figure S16 Normalized cross-correlation (R) maps calculated for residual EBSPs of the dual-phase steel after subtraction of the dominant BCC ferrite/martensite contribution.*

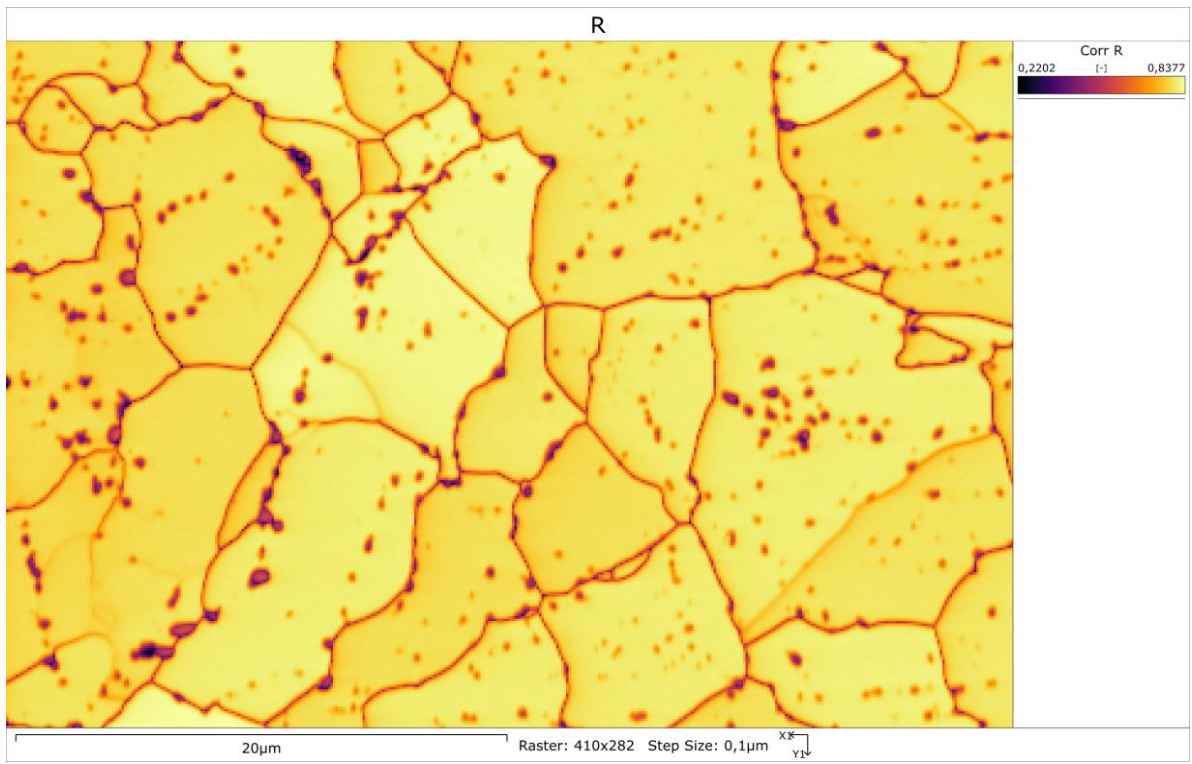

*Figure S17 Normalized cross-correlation (R) maps calculated for experimental EBSPs acquired from the H11 steel prior to residual analysis.*

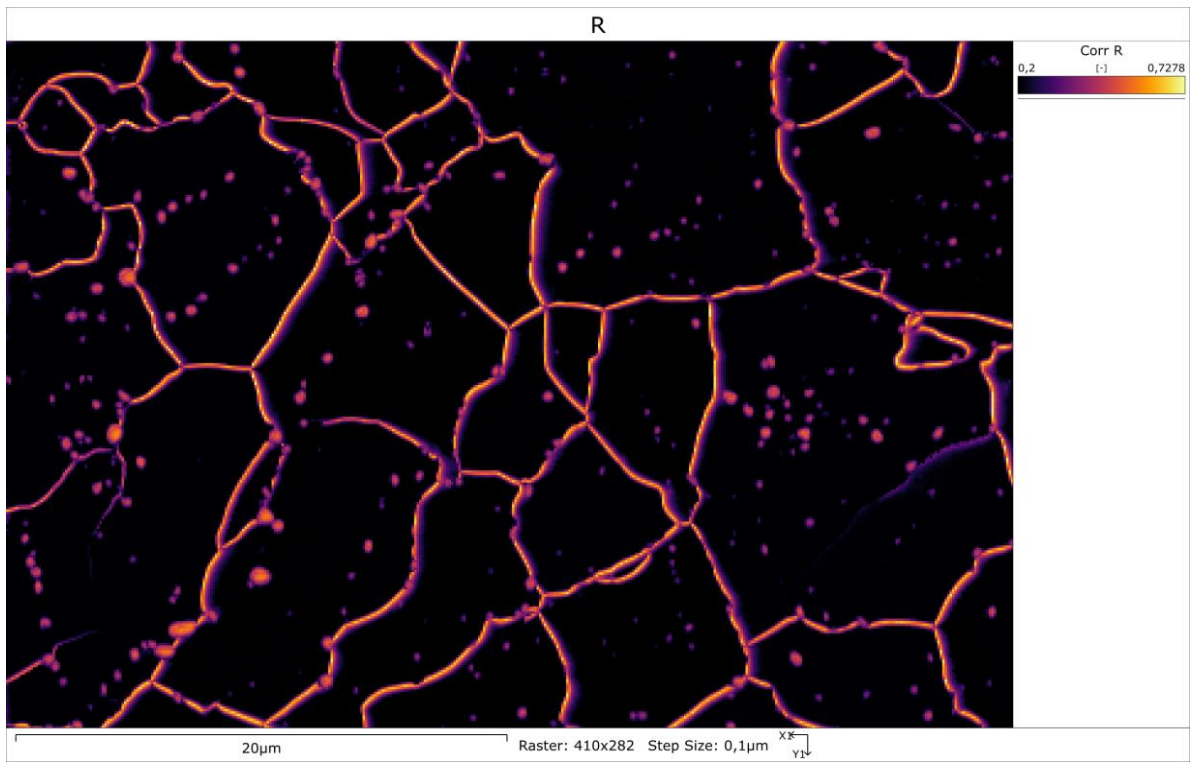

*Figure S18 Normalized cross-correlation (R) maps calculated for residual EBSPs of the H11 steel after subtraction of the dominant BCC ferrite contribution*

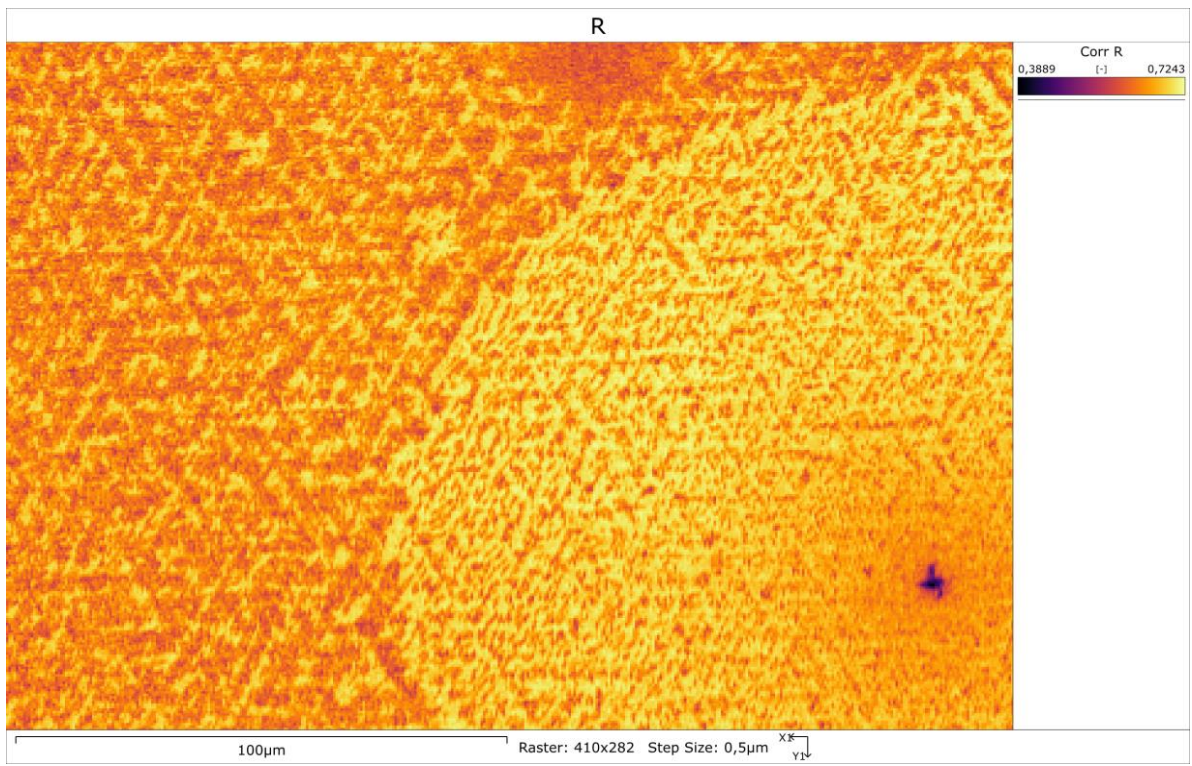

*Figure S19 Normalized cross-correlation (R) maps calculated for experimental EBSPs acquired from the Cu thin film prior to residual analysis*

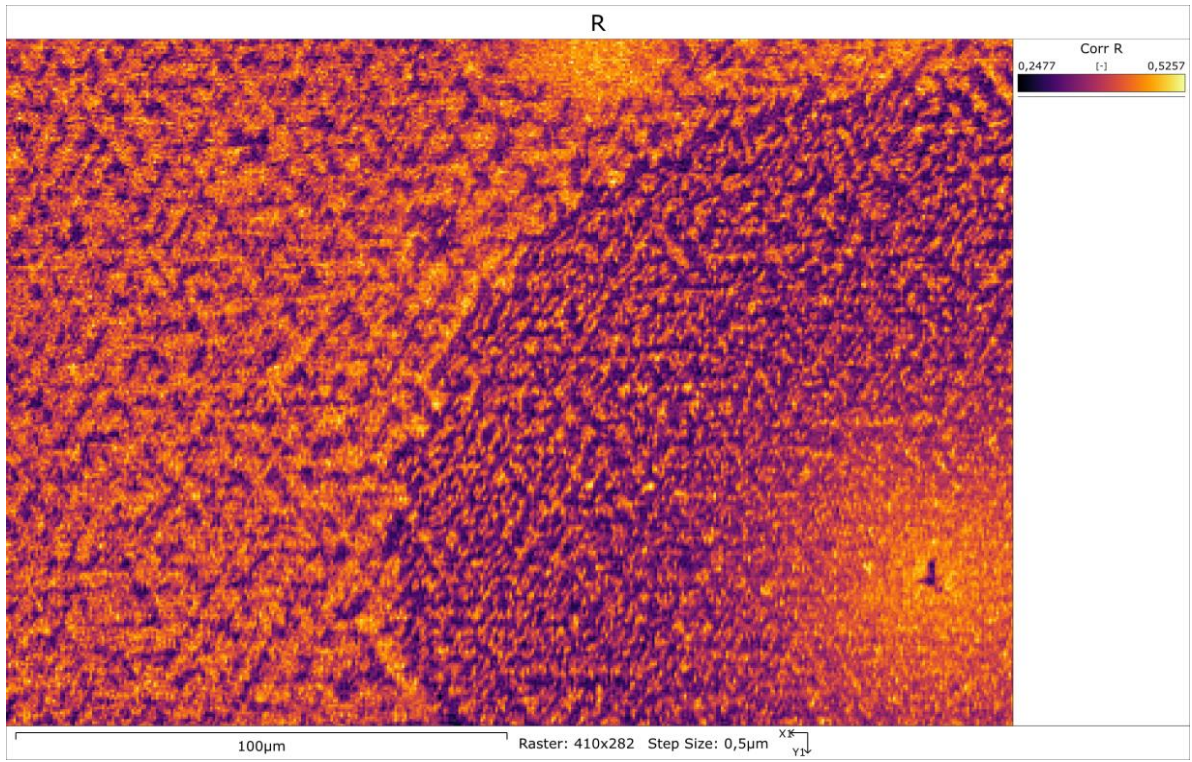

*Figure S20 Normalized cross-correlation (R) maps calculated for residual EBSPs of the Cu thin film after subtraction of the dominant orientation contribution*


\end{document}